\documentclass[aps,pra,twocolumn]{revtex4}

\usepackage{graphicx}
\usepackage{subfigure}
\usepackage{amsmath}
\usepackage{amssymb}
\usepackage{mathrsfs}
\usepackage{hyperref}
\hypersetup{colorlinks=true, linkcolor=black, citecolor=black, urlcolor=black}

\newcommand{\ket}[1]{\left| #1 \right>} 
\newcommand{\bra}[1]{\left< #1 \right|} 
\DeclareMathOperator{\sgn}{sgn}

\begin{document}
\title{Enhanced stripe phases in spin-orbit-coupled Bose-Einstein condensates 
in ring cavities}
\author{Farokh Mivehvar}
\author{David L.~Feder}
\email[Corresponding author: ]{dfeder@ucalgary.ca}
\affiliation{Institute for Quantum Science and Technology, 
Department of Physics and Astronomy,
University of Calgary, Calgary, Alberta, Canada T2N 1N4}
\date{\today}
\begin{abstract}
The coupled dynamics of the atom and photon fields in optical ring cavities
with two counter-propagating modes give rise to both spin-orbit interactions as 
well as long-ranged interactions between atoms of a many-body system. At zero
temperature, the interplay between the two-body and cavity-mediated 
interactions determines the ground state of a Bose-Einstein condensate. In this
work, we find that cavity quantum electrodynamics in the weak-coupling regime
favors a stripe-phase state over a plane-wave phase as the strength of 
cavity-mediated interactions increases. Indeed, the stripe phase is 
energetically stabilized even for 
condensates with attractive intra- and inter-species interactions for 
sufficiently large cavity interactions. The elementary excitation spectra in 
both phases correspond to linear dispersion relation at long wavelengths, 
indicating that both phases exhibit superfluidity, though the plane-wave phase 
also displays a characteristic roton-type feature. 
The results suggest that even in the weak coupling regime cavities can yield 
interesting new physics in ultracold quantum gases.

\end{abstract}
\maketitle

\section{Introduction}
\label{sec:introduction}

The experimental realization of Bose-Einstein 
condensation (BEC) has opened many opportunities for realizing new many-body 
phases~\cite{Anderson-1995,Davis-1995,Bloch-2008}. Ultracold atoms trapped in 
laser-generated optical lattice potentials experience crystalline environments 
and exhibit a variety of intriguing phenomena~\cite{Lewenstein-2007}, most 
notably the superfluid--Mott-insulator phase transition~\cite{Greiner-2002}. 
There are numerous proposals for inducing gauge fields in quantum gases by 
means of laser light~\cite{Dalibard-2011}, and recently abelian~\cite{Lin-2009} 
and non-abelian~\cite{Lin-2011-a} gauge fields have been realized. In the 
latter work an equal combination of Rashba and Dresselhaus spin-orbit (SO) 
couplings were induced via two-photon Raman transitions. These developments 
have set the stage for realizing topological states in these 
systems~\cite{Qi-2011}.  

The single-particle energy dispersion of a SO-coupled atom is a momentum-space
double well, which is two-fold degenerate in the symmetric case~\cite{Lin-2009}.
In a Bose-Einstein condensate (BEC) of atoms, the two-body interactions lift 
this degeneracy and drive the BEC into either a plane wave phase (PWP) or a 
stripe phase (SP), depending on the strength and sign of the intra- and 
inter-species two-body interactions~\cite{Wang-2010,Ho-2011,Li-2012,Zheng-2012}.
In the PWP, all atoms condense into one of the two single-particle energy
minima, while the SP is a superposition state of the minima and the total 
BEC density exhibits faint fringes~\cite{Lu-2013}. Additional phases are found
for fully three-dimensional SO interactions~\cite{Liao2015}. When a SO-coupled 
quantum 
gas is confined in an optical lattice, the ground state of the system exhibits 
a variety of magnetic orderings in the Mott-insulator regime, such as 
ferromagnetic, antiferromagnetic, spin spiral, vortex and antivortex crystals, 
and skyrmion crystal phases~\cite{Cole-2012,Radic-2012,Cai-2012}. The 
superfluid to Mott-insulator phase transition of SO-coupled quantum gases has 
also been investigated~\cite{Cole-2012,Grab-2011}.

In laser-based approaches to generating SO couplings, the radiation field is
treated classically and one ignores the back-action of the atoms on it. 
Confining the radiation field to within an optical cavity leads to a coherent 
exchange of energy and momentum between atoms and photons~\cite{Kimble-1998}. 
The back-action of the atoms on the photon fields is no longer negligible,
leading to complex coupled dynamics of the matter and radiation fields in which 
both entities are affected by one another and must be treated on the same 
footing~\cite{Ritsch-2013}. As a consequence, cavity-mediated long-range 
interactions are induced between atoms, yielding novel collective phenomena in 
atomic systems~\cite{Baumann-2010}. A few schemes have been recently proposed 
to induce SO coupling in ultracold atoms via cavity quantum 
electrodynamics~\cite{Mivehvar-2014,Dong-2014,Deng-2014,Dong2015} and to couple 
a laser-induced SO-coupled BEC to the cavity field~\cite{Padhi-2014}. These 
schemes exhibit a wealth of physics, including strong synthetic magnetic 
fields, a cavity-mediated Hofstadter spectrum, and a variety of magnetic 
orders. 

In this work we investigate the ground state and the elementary excitations of 
a spinor BEC at zero temperature subject to ring-cavity-induced SO 
interactions~\cite{Mivehvar-2014}. Here we consider lossy cavities where a 
steady-state photon population is maintained by the application of external 
pump lasers. The cavity photons mediate infinite-range interactions between 
atoms, whose strengths can be tuned experimentally by adjusting the amplitudes 
of the pump lasers. The sign of these interactions can be made positive or 
negative depending on the cavity detuning, the frequency difference between the 
applied pump lasers and the cavity. These cavity-mediated interactions compete 
with the inherent two-body interactions between atoms to determine the ground 
state of the SO-coupled BEC. In particular, stripe phases are always favored
when positive cavity-mediated interactions dominate the two-body-interactions,
even in the case where the intrinsic atomic interactions (both intra- and 
inter-species) are attractive. Asymmetry in the strength of cavity-mediated 
interactions for different spin components yields stripe-phase states with an 
arbitrary number of atoms in the left or right minimum of the single-particle 
dispersion relation, so that the magnetization varies continuously from zero
in the stripe phase to unity in the plane-wave phase. This behavior allows us 
to identify a novel stripe-phase order parameter, and to identify its
associated mean-field critical exponent.

Consideration of the quantum fluctuations around the mean-field ground states 
reveals that the particle-hole elementary excitation spectra in both PWP and SP 
have the usual linear sound-like dispersion relation at long wavelengths, an 
indication of superfluidity. In the PWP, the dispersion relation also 
exhibits a roton-type feature at the same wave vector that charactizes the 
fringe periodicity in the SP, which could be used experimentally as a 
distinguishing feature. The critical transition between the PWP and SP occurs
when the energy of this minimum falls below zero. Unlike for the PWP, in the 
SP the speed of sound depends strongly on the cavity-mediated interactions. The 
speed of sound is found fall below zero at a critical value of the cavity
interactions and inter-species interactions strength, but this appears to 
signal a phase transition to a phase-separated state. Overall, the ring-cavity 
environment provides an experimentally convenient framework for exploring 
exotic ground states of SO-coupled BECs. 

The manuscript is organized as follows. In Section~\ref{sec:model}, we start 
from the full atom-photon Hamiltonian density for a lossy but pumped cavity, to 
derive an effective atomic Hamiltonian with the photon fields eliminated.
The ground state of this effective Hamiltonian is explored in 
Section~\ref{sec:ground-state} using both a variational method and by solving 
the generalized Gross-Pitaevskii equations. The remainder of this Section is 
devoted to an analysis of the elementary excitations. 
A discussion of the results 
and conclusions are found in Sec.~\ref{sec:conclusions}. 
Appendices~\ref{app:atomic-adiabatic-elimination} 
and \ref{app:cavity-adiabatic-elimination} provide details of the adiabatic 
elimination of the atomic excited state and cavity fields, respectively.        

\section{Model and Hamiltonian}
\label{sec:model}

Consider spin-1 bosonic atoms inside a ring 
cavity with two driven counter-propagating running modes $\hat{A}_1e^{ik_1z}$ 
and $\hat{A}_2e^{-ik_2z}$, where $\hat{A}_j$ is the annihilation operator for 
the photon in $j$th mode with wave vector $k_j=\omega_j/c$ and $z$ is the 
direction along the cavity axis. Without loss of generality, one can assume 
that the wave vectors $k_1$ and $k_2$ of the two modes are approximately equal 
to each other, $k_R\equiv k_1\approx k_2$~\cite{Comment1}. The mode 
$\hat{A}_1e^{ik_Rz}$ ($\hat{A}_2e^{-ik_Rz}$) propagates to the right (left) and 
solely induces the atomic transition $\ket{a}\rightarrow\ket{e}$ 
($\ket{b}\rightarrow\ket{e}$), where $\{\ket{a},\ket{b}\}$ are non-degenerate 
pseudospin states of interest and $\ket{e}$ is an excited state. The two cavity 
modes $\hat{A}_j$ are assumed to be sufficiently populated to justify omitting 
associated degenerate modes $\hat{A}'_j$. In principle, a state-independent 
external potential $V_{\rm ext}(\bf r)$ would need to be imposed to confine 
atoms inside the cavity. The single-particle Hamiltonian density in the dipole 
and rotating-wave approximations is
\begin{align} \label{total-single-atom-H}
\mathcal{H}^{(1)}=\mathcal{H}_{\rm at}^{(1)}+H_{\rm cav}
+\mathcal{H}_{\rm ac}^{(1)},
\end{align}
with
\begin{align*}
\mathcal{H}_{\rm at}^{(1)}&=\left[-\frac{\hbar^2}{2m}\nabla^2
+V_{\rm ext}(\mathbf{r})\right]I_{3\times3}+
\sum_{\tau\in\{a,b,e\}}\varepsilon_\tau \sigma_{\tau \tau},
\nonumber\\
H_{\rm cav}&=\hbar\sum_{j=1,2} \omega_j \hat{A}^\dagger_j \hat{A}_j
+i\hbar \sum_{j=1,2} \left(\eta_j \hat{A}_j^\dagger e^{-i\omega_{{\rm p}j}t} 
- \text{H.c.}\right),
\nonumber\\
\mathcal{H}_{\rm ac}^{(1)}&=\hbar 
\left[\left(\mathscr{G}_{ae}e^{ik_Rz}\hat{A}_1\sigma_{ea}
+\mathscr{G}_{be}e^{-ik_Rz}\hat{A}_2\sigma_{eb}\right) +\text{H.c.} \right],
\end{align*}
where $\varepsilon_\tau$ are the internal atomic-state energies, 
$\sigma_{\tau \tau'}=\ket{\tau}\bra{\tau'}$, and $I_{3 \times 3}$ is the 
identity matrix in the internal atomic-state space. The atom-photon coupling
for the transition $\tau\leftrightarrow\tau'$ is denoted 
$\mathscr{G}_{\tau\tau'}$, and 
H.c.\ stands for the Hermitian conjugate. The cavity mode $\hat{A}_j^\dagger$ 
is driven by a pump laser with frequency $\omega_{{\rm p}j}$ and amplitude 
$\eta_j$, indicated by the second sum in $H_{\rm cav}$. In this work, in order
to simplify the analytical calculations, $V_{\rm ext}(\mathbf{r})$ is set to
zero. In reality, one might imagine a very weak (almost unbound) confining 
potential along the cavity axis $z$ but a standard harmonic trap in the radial 
direction. The details of the transverse confining potential are not important 
for the analysis presented in this work.

After expressing Hamiltonian~\eqref{total-single-atom-H} in the rotating 
frame of the pump lasers~\cite{Maschler-2008} and assuming that the atomic 
detunings $\Delta_1=\omega_1-\varepsilon_{ea}/\hbar$ and 
$\Delta_2=\omega_2-\varepsilon_{eb}/\hbar$ are large compared to 
$\varepsilon_{ba}/\hbar=(\varepsilon_b-\varepsilon_a)/\hbar$, one can 
adiabatically eliminate the atomic excited state to obtain an effective 
Hamiltonian $\mathcal{H}_{\rm SO}'^{(1)}$ for the ground pseudospin states 
$\{\ket{1},\ket{2}\}\equiv\{\ket{b},\ket{a}\}$. 
The details are presented in Appendix \ref{app:atomic-adiabatic-elimination}.
In the limit of a very weak confining potential along the cavity axis 
$\hat{z}$, one can assume that the momentum $p_z=\hbar k_z$ is a good quantum
number. Alternatively one could consider approximately uniform quantum gases in
a box potential where $V_{\rm ext}(\mathbf r)=0$ except at the boundaries; 
such a potential has recently been realized experimentally~\cite{Gaunt-2013}. 
One can then transform to the co-moving frame of the cavity modes by 
applying the unitary transformation $\mathscr{U}_2=e^{-ik_Rz\sigma_z}$ (where 
$\sigma_z=\sigma_{11} - \sigma_{22}$ is the third Pauli matrix, see also 
Appendix \ref{app:atomic-adiabatic-elimination}). The kinetic-energy part of 
the Hamiltonian density $\mathcal{H}_{\rm SO}''^{(1)}\equiv 
\mathscr{U}_2 \mathcal{H}_{\rm SO}'^{(1)} \mathscr{U}_2^\dagger$ 
associated with the momentum $p_z$, Eq.~(\ref{eq:SM-2nd-UT-applied-H}), then
takes the familiar form of an equal Rashba-Dresselhaus SO coupling:
$\frac{1}{2m}(p_zI_{2\times2}+\hbar k_R \sigma_z)^2$,
which is characterized by a double-well energy dispersion~\cite{Lin-2011-a}.

In the presence of dissipation, such as when the decay rate $\kappa$ of both
cavity modes is non-zero, one should in principle numerically solve the 
associated master equation~\cite{Meystre-1999}. That said, in the weak-coupling
regime when $\kappa$ is the dominant energy scale, $\kappa\gg(\mathscr{G}_{ae},
\mathscr{G}_{be})$,
the master equation approach is equivalent to including dissipation in the 
Heisenberg equations of motion for the cavity fields: 
$\partial_t{\hat{A}}_j=-i[\hat{A}_j,\mathcal{H}_{\rm SO}''^{(1)}]/\hbar
-\kappa\hat{A}_j$~\cite{Ritsch-2013}. The cavity fields quickly reach steady 
states, allowing them to be adiabatically eliminated. Setting 
$\partial_t{\hat{A}}_j=0$ one obtains steady-state expressions for $\hat{A}_j$ 
that can be substituted into $\mathcal{H}_{\rm SO}''^{(1)}$ to yield
an effective atomic Hamiltonian; the details are relegated to 
Appendix~\ref{app:cavity-adiabatic-elimination}. 

The resulting effective many-body Hamiltonian reads
\begin{align} \label{eq:totall-eff-H}
H_{\rm eff}&=\int d^3r \left(\hat{\boldsymbol\Psi}^\dagger 
\mathcal{H}_{\rm SO}^{(1)} \hat{\boldsymbol\Psi}
+\frac {1}{2}g_{1} \hat{n}_{1}^2+\frac {1}{2}g_{2} \hat{n}_{2}^2
+g_{12} \hat{n}_{1}\hat{n}_{2} \right)
\nonumber\\
&+\sum_{\tau=1,2} U_{\tau} \hat{N}_\tau^2
+U_{\pm}\hat{S}_+\hat{S}_- + U_{\mp}\hat{S}_-\hat{S}_+
+2U_{\rm ds} \hat{N} \hat{S}_x,
\end{align}
where 
$\hat{\boldsymbol\Psi}(\mathbf r)=
(\hat{\psi}_1(\mathbf r),\hat{\psi}_2(\mathbf r))^\mathsf{T}$ 
are the bosonic field operators obeying the commutation relation 
$[\hat{\psi}_\tau(\mathbf r),\hat{\psi}_{\tau'}^\dagger(\mathbf r')]
=\delta_{\tau,\tau'}\delta(\mathbf{r-r}')$, 
$\hat{N}_{\tau}=\int \hat{n}_{\tau}(\mathbf r)d^3r
=\int \hat{\psi}_{\tau}^\dagger(\mathbf r)\hat{\psi}_\tau(\mathbf r) d^3r$ 
is the total atomic number operator for pseudospin $\tau\in\{1,2\}$,
$\hat{N}=\hat{N}_{1}+\hat{N}_{2}$ is the total atomic number operator,
and the $x$-component of the total spin operator
is defined in a usual way $\hat{S}_x=\frac{1}{2}(\hat{S}_++\hat{S}_-)$
using the collective pseudospin raising and lowering operators
$\hat{S}_+=\hat{S}_-^\dagger
=\int \hat{\psi}_1^\dagger(\mathbf r)\hat{\psi}_2(\mathbf r) d^3r$.
The atoms in this system experience two kinds of interactions, reflected in
the effective Hamiltonian $H_{\rm eff}$: the standard two-body contact 
interactions and the cavity-mediated long-ranged interactions. Here 
$g_\tau\equiv g_{\tau\tau}$ denotes the two-body intra-species interaction 
strength and $g_{12}$ the two-body inter-species interaction strength. The 
strength of the cavity-mediated interactions are found in 
Appendix~\ref{app:cavity-adiabatic-elimination}:
\label{eq:cavity-mediated-interaction-coeff}
\begin{gather}
U_{1(2)}=\frac{4\hbar \mathscr{G}_0^4 \Delta_{\rm c} (\Delta_{\rm c}^2
-3\kappa^2)}{\Delta^2(\Delta_{\rm c}^2+\kappa^2)^3}\eta_{2(1)}^2,
\nonumber\\
U_{\pm(\mp)}=\frac{4\hbar \mathscr{G}_0^4 \Delta_{\rm c}}{\Delta^2(\Delta_{\rm c}^2+\kappa^2)^3}
\left[\Delta_{\rm c}^2-\left(1+2\frac{\eta_{2(1)}^2}{\eta_{1(2)}^2}\right)\kappa^2\right]\eta_{1(2)}^2,
\nonumber\\
U_{\rm ds}=\frac{4\hbar \mathscr{G}_0^4 \Delta_{\rm c}\left(\Delta_{\rm c}^2-3\kappa^2 \right)}
{\Delta^2(\Delta_{\rm c}^2+\kappa^2)^3}\eta_1 \eta_2,
\end{gather}
where $\mathscr{G}_0\equiv\mathscr{G}_{ae}=\mathscr{G}_{be}$,
$\Delta\equiv\Delta_1=\Delta_2$, and $\Delta_c\equiv\omega_{{\rm p}j}-\omega_j$.
The single-particle part of the effective Hamiltonian density has the familiar 
form of the equal Rashba-Dresselhaus SO coupling:
\begin{align} \label{eq:eff-1-particle-SOC-H}
\mathcal{H}_{\rm SO}^{(1)}&=-\frac{\hbar^2}{2m}
\left[\nabla_{\!\bot\!}^2-\left(-i\partial_{z}+k_R\sigma_z\right)^2\right]
+V_{\rm ext}(\mathbf r)\nonumber\\
&+\frac{1}{2}\hbar\delta\sigma_z+\hbar\Omega_R \sigma_x,
\end{align}
with the effective two-photon detuning and Raman coupling given by 
(see Appendix \ref{app:cavity-adiabatic-elimination})
\begin{gather} \label{eq:eff-detuning-Omega}
\delta=\frac{2 \mathscr{G}_0^2 (\Delta_{\rm c}^2-\kappa^2)}
{\Delta(\Delta_{\rm c}^2+\kappa^2)^2}(\eta_2^2-\eta_1^2),
\nonumber\\
\Omega_R=
\frac{2\mathscr{G}_0^2(\Delta_{\rm c}^2-\kappa^2)}{\Delta(\Delta_{\rm c}^2
+\kappa^2)^2}
\left[1-\frac{2\mathscr{G}_0^2 \Delta_{\rm c}}{\Delta(\Delta_{\rm c}^2-\kappa^2)}\right]
\eta_1 \eta_2.
\end{gather}

Before proceeding further, consider briefly some realistic order-of-magnitude 
estimates for various parameters used in the theory based on current 
experiments in ultracold atomic gases and cavity QED. The first experimental 
realization of a synthetic SO coupling was carried out on $^{87}$Rb atoms using 
two counter-propagating Raman laser beams with wavelength $\lambda_R=804.1$~nm 
($E_R=2.33\times10^{-30}$ J)~\cite{Lin-2011-a}; the two-body interaction 
strengths for the desired pseudospin states of $^{87}$Rb atoms are reported to 
be $g_1=5.009\times10^{-51}$ Jm$^3$ and 
$g_2=g_{12}=4.986\times10^{-51}$~Jm$^3$.
With typical average BEC densities $\bar{n}$ of order 
$10^{20}-10^{21}$~m$^{-3}$~\cite{Davis-1995}, one obtains 
$g_\tau\bar{n}/E_R\sim 1$.

One might reasonably expect interesting physics to emerge when the strength of
cavity-mediated interactions becomes comparable to the intrinsic inter-particle
interactions, i.e.\ when $VU_{\tau}/g_{\tau}\sim 1$. Most experimental work is
focused on the strong-cavity limit, where $\mathscr{G}\gg\kappa$; typical 
atom-cavity coupling and cavity decay rates for $^{87}$Rb are 
$\mathscr{G}_{ae}\sim\mathscr{G}_{be}\sim10\kappa
\sim 2\pi\times10$~MHz~\cite{Munstermann2000,Brennecke-2007}. One can attain 
$VU_{\tau}/g_{\tau}\sim 1$ by choosing $\Delta\sim26$~THz, 
$\eta_1=\eta_2=-\Delta_{\rm c}=10$~MHz (for example, $\Delta_{\rm c}\approx
28\kappa$ and $\eta\approx 2.2\kappa$ in Ref.~\onlinecite{Munstermann2000}), 
and a volume $V=10^{-4}$~mm$^3$; for 
these parameters one also obtains $\hbar\Omega_R/E_R\sim4\times10^{-3}$. The
weak coupling regime relevant to the present work can be attained by increasing
the value of $\kappa$, for example by decreasing the reflectivity of the cavity
mirrors. Choosing $\kappa\sim 2\pi\times 100$~MHz one can nevertheless ensure
$VU_\tau/g_1\sim 1$ choosing a larger volume $V=10^{-3}$~mm$^3$ as well as
stronger pump fields and cavity detuning $\eta_1=\eta_2=-3\Delta_{\rm c}=3$~GHz;
these choices yield $\hbar\Omega_R/E_R\sim 4\times10^{-2}$. Further increasing 
the driving field intensities up to $\eta_1=\eta_2=15$~GHz at the fixed 
$\Delta_{\rm c}=-1$~GHz results in cavity-mediated interactions that are an
order of magnitude larger than the two-body interactions $VU_\tau/g_1\sim 30$ 
while $\hbar\Omega_R/E_R\sim 1$. 

In Appendix~\ref{app:cavity-adiabatic-elimination}, which discusses the 
adiabatic elimination of the cavity fields and the origin of the long-ranged 
cavity interactions, quantities such as 
$\mathscr{G}_0^2/[\Delta(\Delta_{\rm c}+i\kappa)]$ and
$\mathscr{G}_0^2N_{\tau}/[\Delta(\Delta_{\rm c}+i\kappa)]$ are assumed to be 
small. Using the weak-coupling values considered above and assuming a typical 
average BEC particle number $N_\tau\sim 10^5$, it is straightforward to verify
that both $\mathscr{G}_0^2/|\Delta(\Delta_{\rm c}+i\kappa)|\ll1$
and $\mathscr{G}_0^2N_{\tau}/|\Delta(\Delta_{\rm c}+i\kappa)|\sim10^{-2}\ll 1$.
Making use of $\kappa\ll\Delta_{\rm c}$ and defining 
$\xi\equiv 2\mathscr{G}_0^2/\Delta\Delta_{\rm c}\ll 1$, one can write
\begin{align}
\label{eq:parametersimplified}
&\quad\Omega_R\approx\frac{\xi\eta_1\eta_2}{\Delta_{\rm c}};\quad
\delta\approx\frac{\xi}{\Delta_c}\left(\eta_2^2-\eta_1^2\right);
\quad U_{\rm ds}\approx\frac{\hbar\xi^2}{\Delta_c}\eta_1\eta_2;\nonumber \\
&U_{1(2)}=U_{\mp(\pm)}\approx\frac{\hbar\xi^2}{\Delta_{\rm c}}\eta_{1(2)}^2.
\end{align}
If $\eta_1=\eta_2$ then $\hbar\delta=0$ and $U_{\rm ds}=U_{1(2)}=U_{\mp(\pm)}$ 
with $U_1/\hbar\Omega_R=\xi\ll 1$. Alternatively, if both pump fields are
non-zero ($\eta_1,\eta_2\neq 0$), then defining $\delta U\equiv U_2-U_1$ one
obtains $\delta U/\hbar\delta=U_{\rm ds}/\hbar\Omega_R=\xi\ll 1$. These 
relations will be important below when choosing parameters for the theoretical 
calculations.

\section{Ground state and excitations: Analytics}
\label{sec:ground-state}

The above analysis indicates that as long as $\eta_1$ and $\eta_2$ are not too
different from one another then $\delta\ll\Omega_R$; in the following we
therefore restrict calculations to $\delta\simeq 0$. The effective 
single-particle Hamiltonian can be diagonalized, and expressed in the form
$H_{\rm SO}^{(1)}=\sum_{\mathbf{k},\lambda=\pm}\epsilon_\lambda(\mathbf{k})
\hat{\varphi}_\lambda^\dagger(\mathbf{k}) \hat{\varphi}_\lambda(\mathbf{k})$ 
with single-particle energy dispersion relation
\begin{subequations}
\begin{align} \label{eq:energy-dispersion}
\tilde{\epsilon}_\pm(\tilde{\mathbf{k}})\equiv\frac{\epsilon_\pm(\mathbf{k})}
{E_R}=\tilde{k}^2+1 \pm\sqrt{4\tilde{k}_z^2+\tilde{\Omega}_R^2},
\end{align}
and spinor eigenstates
\begin{align} \label{eq:spinor-eigenstates}
\boldsymbol\phi_-(\mathbf{k})&=
\begin{pmatrix}
\sin\theta_{\mathbf k} \\
-\cos\theta_{\mathbf k}
\end{pmatrix};\quad
\boldsymbol\phi_+(\mathbf{k})=
\begin{pmatrix}
\cos\theta_{\mathbf k} \\
\sin\theta_{\mathbf k}
\end{pmatrix},
\end{align}
\end{subequations}
where `$+$' and `$-$' designate the upper and lower band, respectively, and
$\sin 2\theta_{\mathbf k}=\tilde\Omega_R/\sqrt{4\tilde{k}_z^2
+\tilde{\Omega}_R^2}$. The unitless parameters $\tilde{\mathbf{k}}
=\mathbf{k}/k_R$ and $\tilde{\Omega}_R=\hbar\Omega_R/E_R$ are defined for
convenience, where $E_R=\hbar^2 k_R^2/2m$ is the recoil energy. Recall that 
using experimentally motivated parameters as discussed toward the end of 
Sec.~\ref{sec:model}, one can choose $\tilde{\Omega}_R\sim\mathcal{O}(1)$. 
Note that in deriving this result we have assumed that the condensate is 
confined in a box potential with negligible occupation of transverse momentum 
states, i.e.\ $\tilde{\mathbf k}=(0,0,\tilde{k}_z)$. In fact, the nature of 
the transverse confinement is not important in the current work; for example,
instead assuming a strong radial oscillator potential 
$V(\rho)=m\omega_{\rho}^2\rho^2/2$ one would simply replace $\tilde{k}^2$ by 
$\tilde{k}_z^2+\hbar\omega_{\rho}/E_R$ under the assumption that the condensate 
occupied the ground state of the radial oscillator. 

The energy dispersion with respect to 
$\tilde{k}_z$ consists of two bands with a band gap of 
$2\tilde\Omega_R$ at the origin $\tilde{\mathbf{k}}=0$. The lower energy band 
$\tilde{\epsilon}_-(\tilde{\mathbf k})$ is a symmetric double well along the 
$\tilde{k}_z$ direction with the two minima located at 
\begin{equation}
\tilde{k}_z=\pm \tilde{k}_0\equiv\pm \sqrt{1-\tilde\Omega_R^2/4},
\label{eq:k0def}
\end{equation}
for $\tilde{\Omega}_R<2$, and it has a single minimum at $\tilde{k}_z=0$ when 
$\tilde\Omega_R>2$ (the minima along the other two directions always occur at 
$\tilde{\mathbf{k}}_{\!\bot\!}=0$). The operators
$\hat{\boldsymbol{\Phi}}(\mathbf{k})=
(\hat{\varphi}_+(\mathbf{k}),\hat{\varphi}_-(\mathbf{k}))^\mathsf{T}$
annihilate a boson at momentum $\mathbf{k}$ in the upper and lower bands 
and are related to the field operators through 
$\hat{\boldsymbol\Psi}(\mathbf r)=\sum_{\mathbf{k},\lambda=\pm}
e^{i\mathbf{k}\cdot\mathbf{r}}
\boldsymbol\phi_\lambda(\mathbf{k}) \hat{\varphi}_\lambda(\mathbf{k})$. 
Note that the laboratory-frame bosonic field operators $\tilde{\boldsymbol\Psi}(\mathbf r)$ (which gives the observable atomic density distribution) are
related to $\hat{\boldsymbol\Psi}(\mathbf r)$ by the 
unity transformation $\mathscr{U}_2$, i.e.\ 
$\tilde{\boldsymbol\Psi}(\mathbf r)
=\mathscr{U}_2^\dagger\hat{\boldsymbol\Psi}(\mathbf r)$.

The single-particle ground state of the symmetric double well (i.e.\ when 
$\tilde{\Omega}_R<2$) is two-fold degenerate; the atom is either in the left 
minimum at $\tilde{\mathbf{k}}=-\tilde{\mathbf{k}}_0=(0,0,-\tilde{k}_0)$ or the 
right minimum at $\tilde{\mathbf{k}}=\tilde{\mathbf{k}}_0=(0,0,\tilde{k}_0)$. 
The non-interacting $N$-particle ground state, when the cavity-mediated
interactions are also absent, is therefore $(N+1)$-fold degenerate 
(any number of pseudospin-up atoms, up to $N$, can reside in the left well).
Nonetheless, the two-body and cavity-mediated interactions compete with each 
other to lift this degeneracy.

\subsection{Variational Approach}
\label{sec:variational-approach}

In order to determine the nature of the ground state,
we assume the following ansatz for the BEC condensate wavefunction,
\begin{align} \label{eq:wavefunction-ansatz}
\begin{bmatrix}
\psi_1\\
\psi_2
\end{bmatrix}=\sqrt{\bar{n}}
\left\{
c_1 e^{-ik_0z}
\begin{bmatrix}
\cos\theta_{\mathbf{k}_0}\\
-\sin\theta_{\mathbf{k}_0}
\end{bmatrix}
+
c_2 e^{ik_0z}
\begin{bmatrix}
\sin\theta_{\mathbf{k}_0}\\
-\cos\theta_{\mathbf{k}_0}
\end{bmatrix}
\right\}
\end{align} 
where $k_0=k_R\tilde{k}_0$ and $\bar{n}=N/V$ is the average particle density,
with $N$ and $V$ being the total particle number and volume, respectively. The 
variational parameters are $c_1$ and $c_2$ with the normalization constraint 
$|c_1|^2+|c_2|^2=1$. Once they are determined, one can find the relevant 
ground-state quantities such as the total density 
$n(\mathbf{r})=|\psi_1(\mathbf{r})|^2+|\psi_2(\mathbf{r})|^2$, and the 
magnetization per particle 
$s_z(\mathbf{r})=[|\psi_1(\mathbf{r})|^2-|\psi_2(\mathbf{r})|^2]/\bar{n}$:
\begin{align} \label{eq:tot-density}
n(\mathbf{r})
=\bar{n}\left[1+2|c_1c_2|\cos(2k_0z+\gamma)\sin2\theta_{\mathbf{k}_0}\right],
\end{align}
\begin{align} \label{eq:sz}
s_z(\mathbf{r})=\left(|c_1|^2-|c_2|^2\right)\cos2\theta_{\mathbf{k}_0},
\end{align}
where $\gamma$ is the relative phase between $c_1$ and $c_2$.
Note that the magnetization $s_z$ is homogeneous while the total density 
$n(\mathbf{r})$ exhibits fringes in the $z$ direction provided that 
$c_1c_2\neq0$. Constraining $\tilde{\Omega}_R<2$, one can write 
$\sin 2\theta_{{\bf k}_0}=\tilde{\Omega}_R/2$ and
$\cos 2\theta_{{\bf k}_0}=\tilde{k}_0$; then these take the simpler form
$n(z)=\bar{n}\left[1+\tilde{\Omega}_R|c_1c_2|\cos(2k_0z+\gamma)\right]$ and 
$s_z=\tilde{k}_0\left(2|c_1|^2-1\right)$.
The energy functional $E[c_1,c_2]=E_0+E_{\rm int}$ is obtained from 
Eq.~\eqref{eq:totall-eff-H} by replacing the field operators $\hat{\psi}_\tau$ 
with the corresponding condensate wavefunctions $\psi_\tau$. This yields
$E_0=-NE_R\tilde{\Omega}_R^2/4$ and 
\begin{widetext} 
\begin{align} \label{eq:Eint-functional}
E_{\rm int}=\frac{N^2|g_1|}{4V} &\Biggl\{\sgn(g_1)+\tilde{g}_2+4\tilde{U}_1
+2\delta\tilde{U}-2\tilde{U}_{\rm ds}\tilde{\Omega}_R
+\left[2\tilde{g}_{12}-\sgn(g_1)-\tilde{g}_2
+4\left(\tilde{U}_{\rm ss}-\tilde{U}_1\right)-2\delta\tilde{U}\right]
\frac{\tilde{\Omega}_R^2}{8}
\nonumber \\
&+\frac{1}{2}\left(|c_1|^2-|c_2|^2\right)\left(4-\tilde{\Omega}_R^2\right)^{1/2}
\big[\sgn(g_1)-\tilde{g}_2-2\delta{U}\big]
\nonumber \\
&-2|c_1c_2|^2\left[\sgn(g_1)+\tilde{g}_2+4\tilde{U}_1+2\delta\tilde{U}
-2\tilde{g}_{12}
-\left(3\sgn(g_1)+3\tilde{g}_2+8\tilde{U}_1+4\delta\tilde{U}-2\tilde{g}_{12}\right)
\frac{\tilde{\Omega}_R^2}{8}\right]
\Biggr\},
\end{align}
\end{widetext}
where the two-body interaction strengths are rescaled by $|g_1|$ (for example 
$\tilde{g}_2=g_2/|g_1|$) and the cavity-mediated interaction strengths are 
rescaled by $|g_1|/V$ (for example $\tilde{U}_1=VU_1/|g_1|$). In the above 
equations we have defined 
$2\tilde{U}_{\rm ss}\equiv\tilde{U}_\pm+\tilde{U}_\mp$
and $\delta\tilde{U}\equiv \tilde{U}_2-\tilde{U}_1$, and 
$\sgn(g_1)=g_1/|g_1|=\pm1$ denotes the sign of $g_1$. Again, recall that 
using experimentally motivated parameters as discussed toward the end of
Sec.~\ref{sec:model}, $\tilde{U}_{1(2)}\sim\tilde{\Omega}_R\sim\mathcal{O}(1)$. 
$E_0$ is 
the single-particle contribution to the energy and is independent of $c_i$, as 
expected. Minimizing $E_{\rm int}$ with respect to $c_i$ determines the ground 
state of the system. The parameters $\tilde{U}_1$ and $\delta\tilde{U}$ 
(or $\tilde{U}_2$) are the only cavity-mediated interaction parameters having 
an effect on the ground state. 

Consider first the simplest case where $\tilde{g}_2=\sgn(g_1)$ and 
$\delta\tilde{U}=0$, so that only that last line of 
Eq.~(\ref{eq:Eint-functional}) contributes to the interaction energy. Then the 
energy is minimized either with $(c_1,c_2)=(1,0)$ or $(0,1)$, or with 
$c_1=c_2=1/\sqrt{2}$ (neglecting relative phases). The first solution set 
corresponds to all atoms condensing in a single minimum of the single-particle 
energy dispersion (i.e.\ a single plane wave with wave vector $-\mathbf{k}_0$ 
or $\mathbf{k}_0$), labeled the plane wave phase (PWP). In the PWP the total 
density is uniform. The magnetization takes the value $s_z=\pm\tilde{k}_0
=\pm(1-\tilde{\Omega}_R^2/4)^{1/2}$, with the upper (lower) sign corresponding 
to $c_1=1$ ($c_1=0$). For small $\tilde{\Omega}_R$ the magnetization approaches
unity. Note that the PWP is twofold degenerate; that is, all atoms 
can condense in the left ($c_1=1$) or right minimum ($c_2=1$). The second 
solution set corresponds to atoms condensing into a superposition state of 
plane waves. It is characterized by the broken translational symmetry and 
the resulting density $n(z)=n[1+\frac{1}{2}\tilde{\Omega}_R
\cos(2k_0z+\gamma)]$ exhibits spatial variations in the $z$ (i.e.\ 
SO-coupling) direction, so this is referred to as the stripe phase (SP). In
this phase the density oscillations have greatest contrast for large
$\tilde{\Omega}_R\to 2$. The SP magnetization $s_z$ is zero.

The SP solution yields a lower energy than the PWP solution when term in square 
brackets in the last line of Eq.~\eqref{eq:Eint-functional} is positive. 
(Recall $\tilde{g}_2=\sgn(g_1)$ and $\delta\tilde{U}=0$ so that the middle 
line vanishes identically.) The cavity interaction strength that favors the SP 
solution is therefore $\tilde{U}_1>\tilde{U}_{1\rm c}^0$, where 
\begin{equation}
\tilde{U}_{1\rm c}^0\equiv\frac{8\left[\tilde{g}_{12}-\sgn(g_1)\right]
-\left[\tilde{g}_{12}-3\sgn(g_1)\right]\tilde{\Omega}_R^2
}{4(4-\tilde{\Omega}_R^2)}, 
\label{eq:SP-PWP-trans-VA}
\end{equation}
is the critical cavity interaction for the SP-PWP transition.
In the limit of small $\tilde{\Omega}_R$, this becomes 
$\tilde{U}_{1\rm c}^0\simeq\frac{1}{2}[\tilde{g}_{12}-\sgn(g_1)]
+\frac{1}{16}[\tilde{g}_{12}+\sgn(g_1)]\tilde{\Omega}_R^2$. 
If $\tilde{g}_{12}=\sgn(g_1)$ the SP is favored for any non-zero, positive 
cavity interaction in the limit $\tilde{\Omega}_R\rightarrow0$.
In the other hand when $\tilde{\Omega}_R\rightarrow2$ 
and $\tilde{g}_{12}\neq-\sgn(g_1)$, 
the critical cavity interaction $\tilde{U}_{1\rm c}^0$ diverges and SP 
is only favored for very large positive cavity interaction.  

It is important to verify that the total interaction energy, 
Eq.~(\ref{eq:Eint-functional}), remains positive; the system is stable only if
$\partial^2 E_{\rm int}/\partial N^2>0$. Let us examine this first in the SP 
where $c_1=c_2=1/\sqrt{2}$, for a special case where 
$\tilde{U}_{\rm ds}=\tilde{U}_{\rm ss}=\tilde{U}_1$ (and 
$\tilde{g}_2=\sgn(g_1)$ and $\delta\tilde{U}=0$ as before). One obtains
\begin{align} \label{eq:Eint-functional-special-case-SP}
E_{\rm int}=\frac{N^2|g_1|}{4V}
&\bigg\{\frac{1}{8}\left[\tilde{g}_{12}+\sgn(g_1)\right]\left(8+\tilde{\Omega}_R^2\right)\nonumber\\
&+\frac{1}{2}\tilde{U}_1\left(2-\tilde{\Omega}_R\right)^2\bigg\}.
\end{align}
Surprisingly, the SP is energetically stable for two-component attractive BECs 
in the presence of spin-orbit interactions as long as the inter-species 
interaction strength is sufficiently large and positive. 
Substituting the critical cavity interaction $\tilde{U}_{1\rm c}^0$ into 
Eq.~\eqref{eq:Eint-functional-special-case-SP} yields the constraint
\begin{align} \label{eq:g12-sign_g1-Omega-const}
\tilde{g}_{12}\geqslant\sgn(g_1)\frac{\tilde{\Omega}_R\left[(2-\tilde{\Omega}_R)^2-12\right]}
{\tilde{\Omega}_R^3+16}.
\end{align}

In the limit of $\tilde{\Omega}_R\rightarrow0$, for the lowest possible values
of the cavity interaction favoring the SP phase
$\tilde{U}_1\gtrsim\tilde{U}_{1\rm c}^0=\frac{1}{2}[\tilde{g}_{12}-\sgn(g_1)]$,
the SP is energetically stable as long as $\tilde{g}_{12}\geqslant0$, with no 
constraint on the sign of the intra-species interaction strength. Thus, the
infinite-range cavity-mediated atom-atom interactions stabilize attractive
two-component BECs against collapse, even in the absence of a confining 
potential. For larger values of $\tilde{U}_1$ even the inter-species 
interactions can be attractive.

The coefficient of $\tilde{U}_1$ in 
Eq.~(\ref{eq:Eint-functional-special-case-SP}) is strictly positive. Therefore, 
for a given 
parameter set $\{\sgn(g_1),\tilde{g}_{12},\tilde{\Omega}_R\}$ one can choose
arbitrary large positive values of the cavity interaction strength to strongly 
favor SP without compromising stability (i.e.~to satisfy
$\tilde{U}_1>\tilde{U}_{1\rm c}^0$ while ensuring that $E_{\rm int}\geqslant0$).
In other words, the minimal cavity interaction $\tilde{U}_1$ which favors 
a stable SP satisfies
\begin{align}
\tilde{U}_1>{\rm max}\left\{-\frac{\left[\tilde{g}_{12}+\sgn(g_1)\right]\left(8+\tilde{\Omega}_R^2\right)}
{4\left(2-\tilde{\Omega}_R\right)^2},\tilde{U}_{1\rm c}^0\right\}.
\end{align}

The stability of PWP can be investigated in a similar manner. The plane wave 
phase is favored when $\tilde{U}_1<\tilde{U}_{1\rm c}^0$. The positivity 
constraint of the interaction energy in the PWP
\begin{align} \label{eq:Eint-functional-special-case-PWP}
E_{\rm int}=\frac{N^2|g_1|}{2V}
&\bigg\{\sgn(g_1)+\frac{1}{8}\left[\tilde{g}_{12}-\sgn(g_1)\right]\tilde{\Omega}_R^2\nonumber\\
&+\tilde{U}_1\left(2-\tilde{\Omega}_R\right)\bigg\}>0,
\end{align}
imposes a lower bound in the cavity interaction
\begin{align}  
-\frac{8\sgn(g_1)+\left[\tilde{g}_{12}-\sgn(g_1)\right]\tilde{\Omega}_R^2}
{8\left(2-\tilde{\Omega}_R\right)}
<\tilde{U}_1<\tilde{U}_{1\rm c}^0,
\end{align}
beyond which PWP is unstable. Thus, even the PWP becomes energetically stable
for attractive spin-orbit coupled two-component BECs if the cavity-mediated 
interactions are judiciously chosen. 

\begin{figure}[t]
\centering
\subfigure[]{
\includegraphics [width=0.232\textwidth]{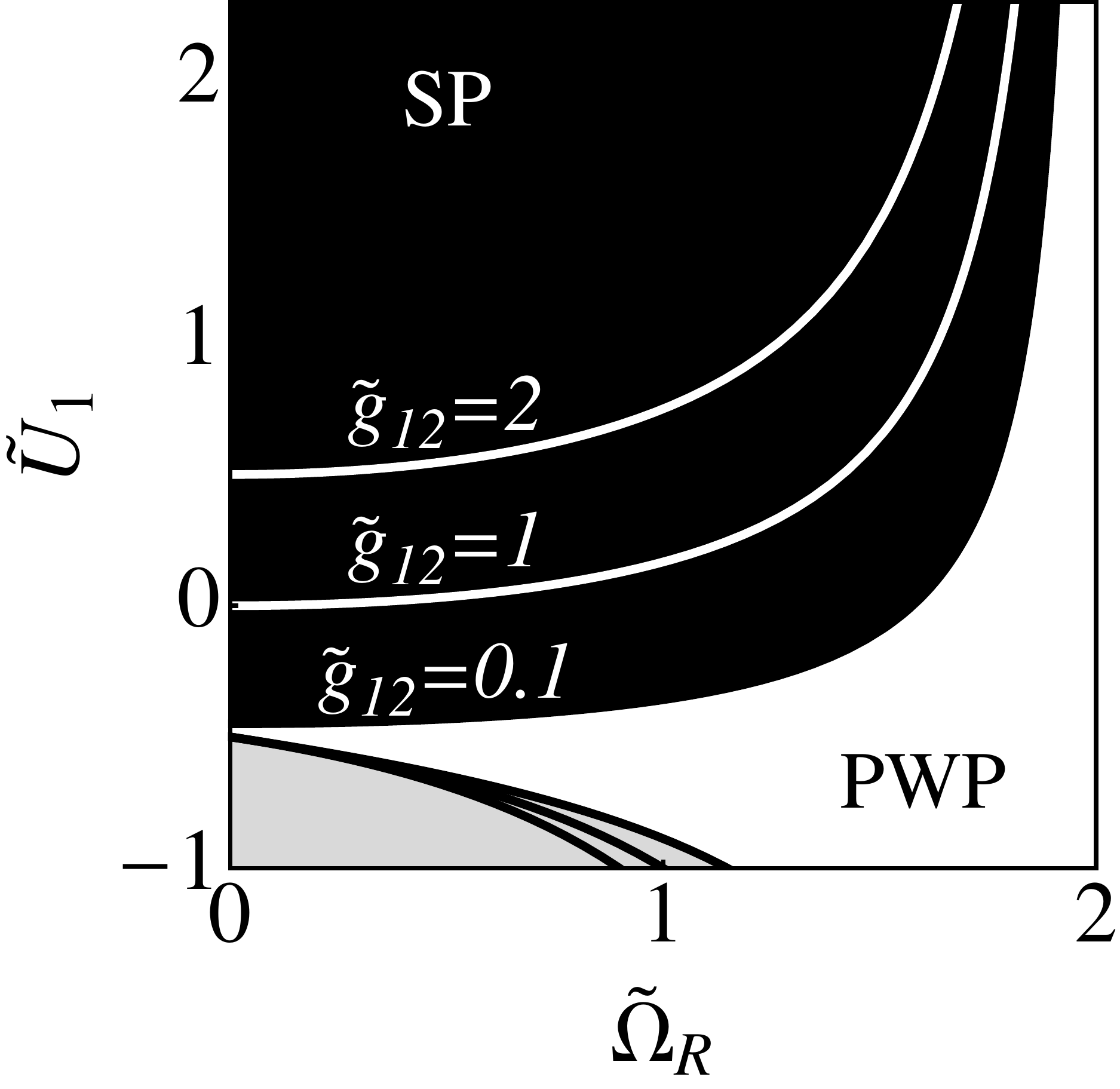}
\label{fig:U-omega-plane}}
\subfigure[]{
\includegraphics [width=0.218\textwidth]{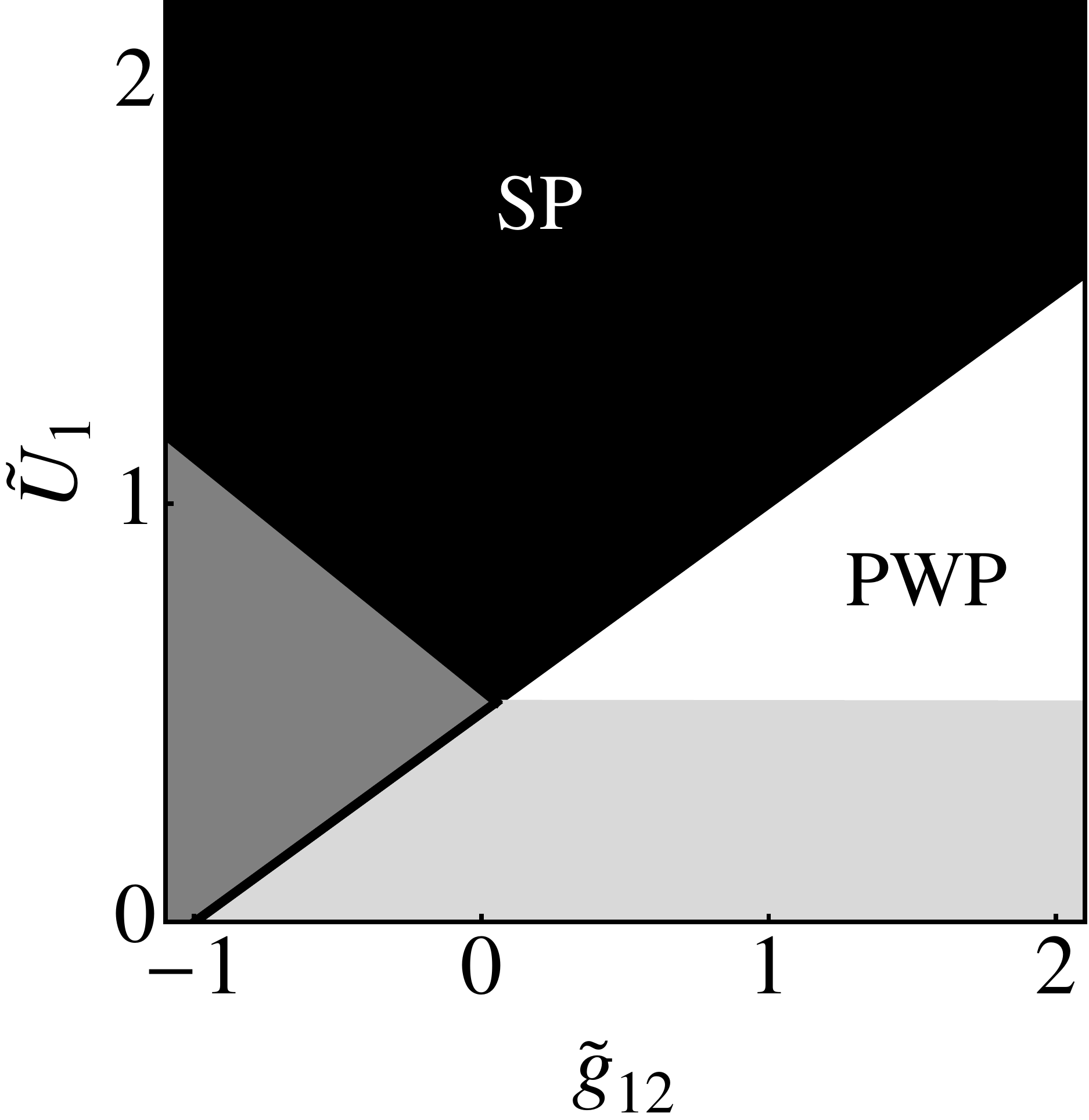}
\label{fig:U-g-plane}}
\caption{Phase diagrams in the (a) $\{\tilde{U}_1,\tilde{\Omega}_R\}$ 
and (b) $\{\tilde{U}_1,\tilde{g}_{12}\}$ parameter planes. The stripe and 
plane-wave phases are denoted by back and white, respectively; dark (light)
grey indicates the regions where the SP (PWP) is unstable. (a) Phase diagram 
for $\sgn(g_1)=\tilde{g}_2=1$ and different values of $\tilde{g}_{12}=0.1$, 1, 
and 2. (b) Phase diagram for $\sgn(g_1)=\tilde{g}_2=-1$ and 
$\tilde{\Omega}_R=0.1$.}
\label{fig:phase-diag-1}
\end{figure}

Figure~\ref{fig:phase-diag-1} depicts the phase diagrams in the 
$\{\tilde{U}_1,\tilde{\Omega}_R\}$ and $\{\tilde{U}_1,\tilde{g}_{12}\}$ 
parameter planes. The phase diagrams are comprised of two physical regions: the
SP and PWP, denoted by black and white in Fig.~\ref{fig:phase-diag-1}, 
respectively. The dark (light) grey indicates the regions where the SP (PWP) is 
energetically unstable. Figure \ref{fig:U-omega-plane} shows the phase diagram 
in the 
$\{\tilde{U}_1,\tilde{\Omega}_R\}$ parameter space for 
$\sgn(g_1)=\tilde{g}_2=1$ and different values of 
$\tilde{g}_{12}$. The stripe phase is favored over an ever-larger parameter space 
as $\tilde{U}_1$ increases as long as $|\tilde{\Omega}_R|<2$ to assure the existence
of a double-well single-particle dispersion. This general 
trend is also evident from Fig.~\ref{fig:U-g-plane}, the phase diagram 
in the $\{\tilde{U}_1,\tilde{g}_{12}\}$ parameter plane for
$\sgn(g_1)=\tilde{g}_2=-1$ and constant $\tilde{\Omega}_R=0.1$, 
where Eq.~(\ref{eq:SP-PWP-trans-VA}) reveals that the 
phase boundary is linear in $\tilde{g}_{12}$ for fixed $\tilde{\Omega}_R$. 

\begin{figure}[t]
\centering
\includegraphics [width=0.45\textwidth]{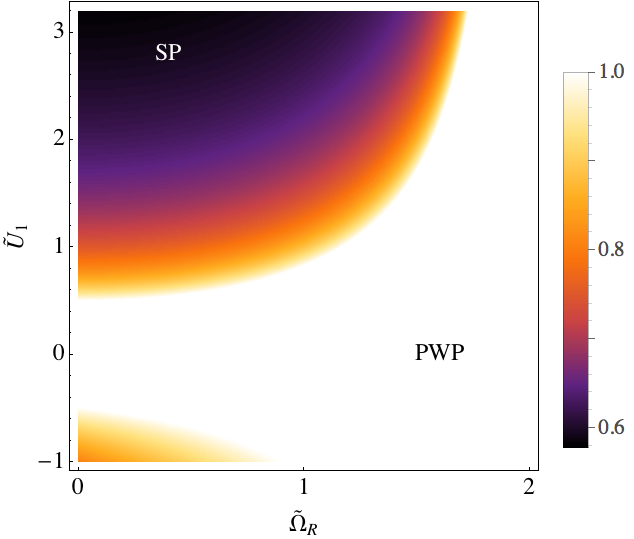}
\caption{(Color online) Density plot of $|c_1|^2$ in the 
$\{\tilde{U}_{1},\tilde{\Omega}_R\}$ parameter plane for 
$\sgn(g_1)=\tilde{g}_2=\delta\tilde{U}=1$, and $\tilde{g}_{12}=2$.
The plane-wave phase begins to be unstable in the left bottom corner.}
\label{fig:g-omega-plane-2}
\end{figure}

Relaxing the constraint considered above that $\delta\tilde{U}=0$ in 
Eq.~(\ref{eq:Eint-functional}), one can prepare any arbitrary superposition 
state, i.e.\ arbitrary $c_1$ and $c_2$ subject to $|c_1|^2+|c_2|^2=1$. 
The plane-wave phase is no longer degenerate; rather, the minimum favored 
depends on the sign of $\delta\tilde{U}$. Figure~\ref{fig:g-omega-plane-2} 
shows the dependence of $|c_1|^2$ in the $\{\tilde{U}_{12},\tilde{\Omega}_R\}$ 
plane for $\sgn(g_1)=\tilde{g}_2=\delta\tilde{U}=1$, and $\tilde{g}_{12}=2$. 
Under these conditons the SP with $|c_1|=|c_2|$ is found only for very large 
$\tilde{U}_{1}\gg\tilde{U}_{1\rm c}$, i.e.\ far from the SP-PWP phase 
boundary $\tilde{U}_{1\rm c}$. Whereas for $\tilde{U}_{1}\rightarrow\tilde{U}_{1\rm c}^{+}$, $|c_1|$ 
increases monotonically until the PWP with $|c_1|^2=1$ is attained for 
$\tilde{U}_{1}<\tilde{U}_{1\rm c}$ (note that the critical value
$\tilde{U}_{1\rm c}\simeq\tilde{U}_{1\rm c}^0$ and is weakly 
dependent on $\delta\tilde{U}$, as discussed below). The 
plane-wave phase begins to be unstable in the left bottom corner of this 
figure. 

\begin{figure}[t]
\centering
\includegraphics [width=0.45\textwidth]{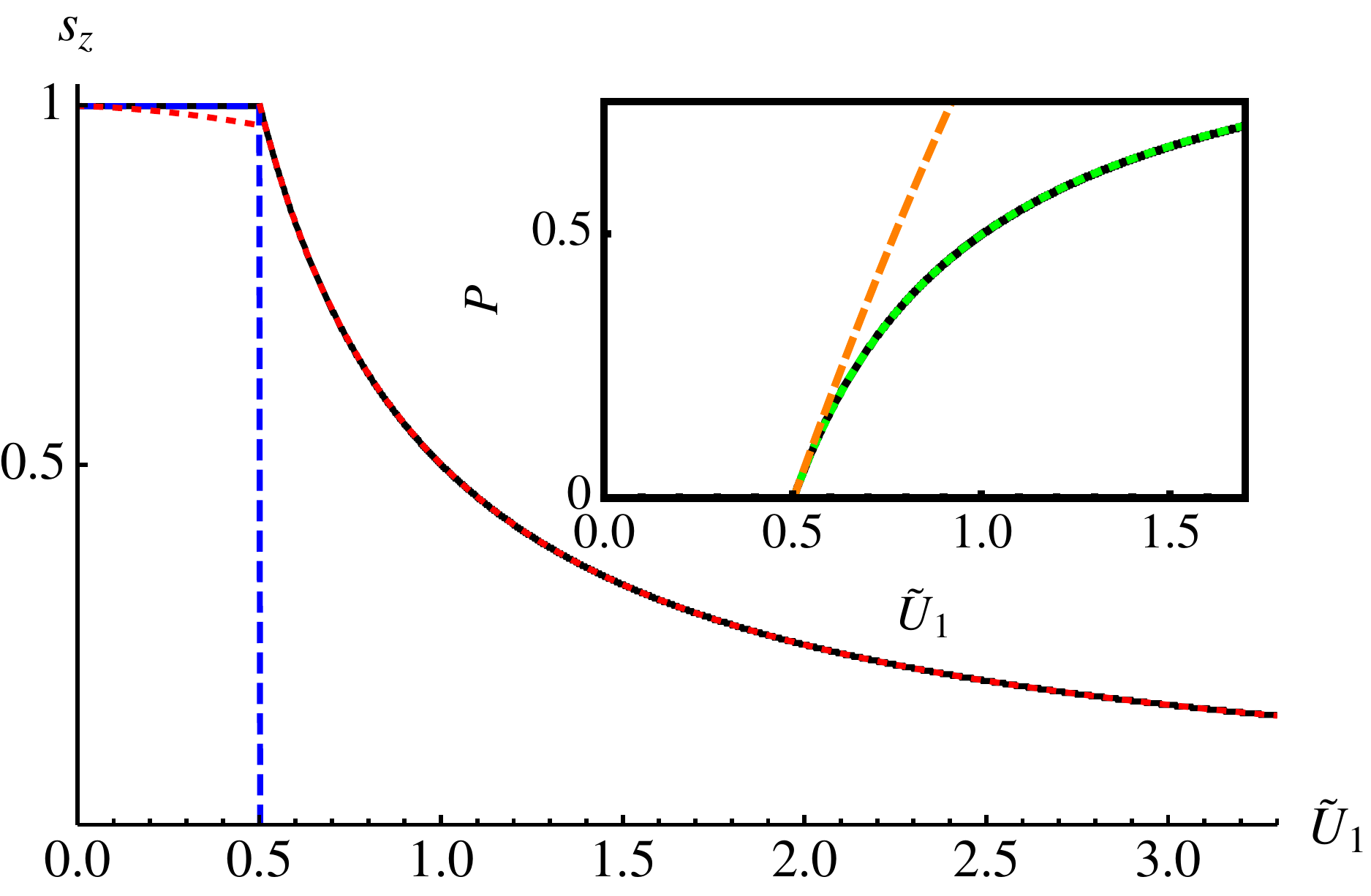}
\caption{(Color online) The magnetization $s_z$ as a function of $\tilde{U}_1$ 
shown as the black solid curve for $\sgn(g_1)=\tilde{g}_2=\delta\tilde{U}=1$, 
$\tilde{g}_{12}=2$, and $\tilde{\Omega}_R=0.1$. The blue dashed curve 
represents the magnetization when $\delta\tilde{U}=0$. The red dotted curves 
are the magnetization computed from solutions of the coupled 
Gross-Pitaevskii equations in the SP and PWP assuming $\tilde{U}_{\rm ss}
=\tilde{U}_{\rm ds}=\tilde{U}_1$ for the same parameters as the solid black 
curve, and $|g_1|\bar{n}/E_R=1$. Inset: the SP order parameter $P$ is shown as a
function of $\tilde{U}_1$ (black curve); an analytical approximation (dashed
green curve) and the behavior near the critical point (orange dashed curve) are
shown for comparison.}
\label{fig:sz}
\end{figure}

The magnetization $s_z=\tilde{k}_0\left(2|c_1|^2-1\right)$ as a function of 
$\tilde{U}_1$ is illustrated with the black solid curve in Fig.~\ref{fig:sz} 
for $\sgn(g_1)=\tilde{g}_2=\delta\tilde{U}=1$, $\tilde{g}_{12}=2$, and 
$\tilde{\Omega}_R=0.1$. For contrast, the magnetization when 
$\delta\tilde{U}=0$ is also shown (blue dashed curve). Note that while the 
sign of the magnetization in the PWP is arbitrary for the $\delta\tilde{U}=0$
case (a spontaneously broken symmetry in the ground state), in the present case 
the sign of $s_z$ always follows that of $\delta\tilde{U}.$ On the PWP side,
the magnetization is fixed at its maximal value $s_z=\tilde{k}_0$; for 
$\tilde{U}_1\gtrsim\tilde{U}_{1\rm c}$ on the SP side, the magnetization 
decreases sharply before reaching an asymptotic value deep within the SP phase.

For small $\delta\tilde{U}$ and $\tilde{\Omega}_R$, the SP-PWP phase 
transition occurs at almost the same value of the critical cavity interaction 
$\tilde{U}_{1\rm c}^0=0.5$ obtained using Eq.~(\ref{eq:SP-PWP-trans-VA}) which 
assumed $\delta\tilde{U}=0$. Near the phase transition point on the SP side,
one can write $c_1=1-x^2$ and $c_2=\sqrt{2}x$, where $x\ll1$ and 
$c_1^2+c_2^2\simeq1+\mathcal{O}(x^4)$. 
Setting the term proportional to $x^2$ in
$E_{\rm int}[c_1=1,c_2=0]-E_{\rm int}[c_1=1-x^2,c_2=\sqrt{2}x]$
equal to zero yields a modified critical cavity interaction
\begin{align}
\label{eq:Ucbetter}
\tilde{U}_{1\rm c}=\tilde{U}_{1\rm c}^0
-\left[\frac{2-(4-\tilde{\Omega}_R^2)^{1/2}-\frac{1}{2}\tilde{\Omega}_R^2}
{4-\tilde{\Omega}_R^2}\right]\delta\tilde{U}.
\end{align}
In the small $\tilde{\Omega}_R$ limit this may be simplified to
$\tilde{U}_{1\rm c}\simeq\frac{1}{2}[\tilde{g}_{12}-\sgn(g_1)]
+\frac{1}{16}[\tilde{g}_{12}+\sgn(g_1)+\delta\tilde{U}]\tilde{\Omega}_R^2$,
which is the same critical cavity interaction $\tilde{U}_{1\rm c}^0$ obtained 
above in the small $\tilde{\Omega}_R$ limit, save for the 
$\delta\tilde{U}$-dependent correction.

The behavior of the magnetization for $\tilde{U}_1>\tilde{U}_{1\rm c}$ suggests 
that one can define the order parameter for the stripe phase to be 
$P=1-s_z/\tilde{k}_0=2(1-c_1^2)$. As desired, this vanishes in the
PWP (here we only consider a PWP with momentum $-\mathbf{k}_0$) and takes a
nonzero value in SP. The order parameter is shown in the inset of 
Fig.~\ref{fig:sz}. The discontinuity in the derivative of $P$ with 
$\tilde{U}_1$ suggests that the SP-PWP quantum (zero-temperature) phase 
transition is second order. It is therefore of interest to determine the 
(mean-field) exponent $\beta$ for the order parameter $P$ in the vicinity of
the transition point. Substituting $\tilde{U}_1=\tilde{U}_{1\rm c}+\chi$ into 
the energy functional $E_{\rm int}$ and minimizing it with respect to $c_1$ 
yields
\begin{align} \label{c1-MF}
c_1=\sqrt{\frac{2\delta\tilde{U}\left(4-\tilde{\Omega}_R^2\right)^{1/2}+\chi\left(4-\tilde{\Omega}_R^2\right)}
{2\delta\tilde{U}\left(4-\tilde{\Omega}_R^2\right)^{1/2}+2\chi\left(4-\tilde{\Omega}_R^2\right)}}. 
\end{align}
The order parameter $P=2(1-c_1^2)$ computed using this expression for $c_1$ is 
illustrated as the green dashed curve in the the inset of Fig.~\ref{fig:sz},
and is in excellent agreement with the numerical results of the variational 
approach, shown as the black solid curve. Taylor expanding $c_1$ in 
Eq.~(\ref{c1-MF}) for small $\chi$ and $\tilde{\Omega}_R$ up to first and 
second order, respectively, one obtains 
$c_1^{\rm MF}\simeq1-\chi/2\delta\tilde{U}$ 
(the term proportional to $\chi\tilde{\Omega}_R^2$ is also omitted).
This yields the mean-field order parameter $P_{\rm MF}=2\chi/\delta\tilde{U}
=2(\tilde{U}_1-\tilde{U}_{1\rm c})^\beta/\delta\tilde{U}$ and a critical 
exponent $\beta=1$. The behavior of the order parameter near the transition
point fits well to $P$, as is shown by the orange dashed curve in the inset of 
Fig.~\ref{fig:sz}.

In principle, it is not valid to consider $\delta\tilde{U}\neq 0$ while 
at the same time assuming that $\tilde{\delta}\equiv\hbar\delta/E_R=0$. Rather,
if $\eta_1\neq\eta_2\neq 0$ but $\eta_1\sim\eta_2$, then
Eqs.~(\ref{eq:parametersimplified}) state that 
$\tilde{\delta}\sim\delta\tilde{U}$ whenever $\tilde{U}_1\sim\tilde{\Omega}_R$.
That said, in Fig.~\ref{fig:sz} the parameters are chosen so that 
$\tilde{\Omega}_R=0.1\ll\delta\tilde{U}=1$. One can therefore expect
$\tilde{\delta}\ll\delta\tilde{U}$ by a similar ratio, which again justifies
neglecting it. 

Consider briefly the effect of keeping a non-zero but small value of 
$\tilde{\delta}$. The single-particle dispersions of the spin-orbit 
Hamiltonian~(\ref{eq:eff-1-particle-SOC-H}) become
\begin{equation}
\tilde{\epsilon}_{\pm}(\tilde{\bf k})=\tilde{k}_z^2+1
\pm\sqrt{\frac{1}{4}\left(4\tilde{k}_z+\tilde{\delta}\right)^2
+\tilde{\Omega}_R^2}
\end{equation}
rather than the expressions given in Eq.~(\ref{eq:energy-dispersion}). The 
associated (orthogonal) eigenvectors have the same form as 
Eqs.~(\ref{eq:spinor-eigenstates}) but now 
$\sin 2\theta_{\mathbf k}=\tilde{\Omega}_R/\sqrt{
\frac{1}{4}\left(4\tilde{k}_z+\tilde{\delta}\right)^2+\tilde{\Omega}_R^2}$. For 
$\tilde{\delta}\neq 0$, the lower double-well dispersion curve 
$\tilde{\epsilon}_-$ is no longer symmetric; rather, the right well is lower
(higher) when $\tilde{\delta}>0$ ($\tilde{\delta}<0$). Thus, in the absence of
particle interactions a PWP is energetically favored in one well or the other
with no ambiguity. The presence of $\tilde{\delta}$ precludes a simple form 
like Eq.~(\ref{eq:k0def}) for the location of the energy minima, but in the 
limit when both $\tilde{\Omega}_R\ll 1$ and $\tilde{\delta}\ll 1$ one obtains
\begin{equation}
\tilde{k}_0\approx 1-\frac{\tilde{\Omega}_R^2}{8}\left(1
-\frac{\tilde{\delta}}{2}\right).
\end{equation}
The lowest-order contribution of $\tilde{\delta}$ is a correction to the 
coefficient of the already small $\tilde{\Omega}_R$-dependent term, and 
therefore the value of $\tilde{k}_0$ is well-approximated by assuming 
$\tilde{\delta}=0$. Likewise, the BEC approximation consists of $\tilde{k}_z$ 
with $\tilde{k}_0$; because $4\tilde{k}_z\to 4\tilde{k}_0\approx 4
\gg\tilde{\delta}$ in the expressions for the single-particle energies and 
eigenvectors above, $\tilde{\delta}$ can be similarly neglected in the 
calculations.

\subsection{Coupled Gross-Pitaevskii Equations}
\label{sec:GGPE}

While the variational calculation discussed in the previous section has 
revealed that a ring cavity can stabilize stripe phases in interacting 
spin-orbit coupled Bose-Einstein condensates, it is important to verify the
results using a more rigorous approach. In this section, the coupled 
Gross-Pitaevskii (GP) equations are derived for both PWP and SP ans\" atze
and the ground state properties are obtained from their solutions. 

\subsubsection{Plane wave phase}
\label{sec:PWP}

The GP equations can be obtained directly from the many-particle
Hamiltonian~\eqref{eq:totall-eff-H}:
\begin{align} \label{eq:GGPEs}
&\left[\frac{\hbar^2}{2m}\hat\triangle_1
+g_1|\psi_1|^2+g_{12}|\psi_2|^2+2U_1N_1+2U_{\rm ds} S_x\right]\psi_1\nonumber\\
&+\big[\hbar\Omega_R+2U_{\rm ss}S_-+U_{\rm ds}N\big]\psi_2
=\mu \psi_1,\nonumber\\
&\left[\frac{\hbar^2}{2m}\hat\triangle_2
+g_2|\psi_2|^2+g_{12}|\psi_1|^2+2U_2N_2+2U_{\rm ds} S_x\right]\psi_2\nonumber\\
&+\big[\hbar\Omega_R+2U_{\rm ss}S_++U_{\rm ds}N\big]\psi_1
=\mu \psi_2,
\end{align}
where $\hat\triangle_1=-\nabla_{\!\bot\!}^2+(-i\partial_{z}+k_R)^2$ and
$\hat\triangle_2=-\nabla_{\!\bot\!}^2+(-i\partial_{z}-k_R)^2$ and the 
BEC wavefunctions for the two spin components are denoted by $\psi_{1(2)}$
rather than $\psi_{1(2)}({\bf r})$ to save space. These equations can be 
simplified in the plane-wave phase by assuming homogeneous wavefunctions 
$\psi_\tau(\mathbf{r})=e^{\pm ik_0z}\bar{\psi}_\tau$, where the upper (lower) 
sign corresponds to a condensate in the right (left) minimum. The GP equations 
are then recast as
\begin{align} \label{eq:GGPEs-PWP}
&\left[\tilde{\mu}-(\tilde{k}_0\pm1)^2\right]\bar{\psi}_1
-\tilde{\Omega}_R\bar{\psi}_2\nonumber \\
&=\frac{|g_1|}{E_R}\bigg\{\left[\left(\sgn(g_1)+2\tilde{U}_1\right)|\bar{\psi}_1|^2
+\left(\tilde{g}_{12}+2\tilde{U}_{\rm ss}\right)|\bar{\psi}_2|^2\right]\bar{\psi}_1
\nonumber \\
&+\tilde{U}_{\rm ds}\left(2|\bar{\psi}_1|^2
+|\bar{\psi}_2|^2\right)\bar{\psi}_2+\tilde{U}_{\rm ds}
\bar{\psi}_1^2\bar{\psi}_2^*\bigg\};\nonumber\\
&\left[\tilde{\mu}-(\tilde{k}_0\mp1)^2\right]\bar{\psi}_2
-\tilde{\Omega}_R\bar{\psi}_1\nonumber \\
&=\frac{|g_1|}{E_R}\bigg\{\left[\left(\tilde{g}_2+2\tilde{U}_2\right)|\bar{\psi}_2|^2
+\left(\tilde{g}_{12}+2\tilde{U}_{\rm ss}\right)|\bar{\psi}_1|^2\right]\bar{\psi}_2
\nonumber \\
&+\tilde{U}_{\rm ds}\left(|\bar{\psi}_1|^2
+2|\bar{\psi}_2|^2\right)\bar{\psi}_1+\tilde{U}_{\rm ds}
\bar{\psi}_2^2\bar{\psi}_1^*\bigg\},
\end{align}
where again the upper (lower) sign in each equation corresponds to a condensate 
in the right (left) minimum, and the chemical potential is expressed in recoil
energy units, $\tilde{\mu}\equiv\mu/E_R$. 

The chemical potential can be obtained from the first of 
Eqs.~(\ref{eq:GGPEs-PWP}) and then substituted into the second. Under the
assumption that both condensate wavefunctions are real, $\tilde{U}_{1}
=\tilde{U}_{\rm ss}=\tilde{U}_{\rm ds}$, and $\sgn(g_1)=\tilde{g}_2$, one 
obtains
\begin{align}
&\frac{|g_1|}{E_R}\left[\big(\tilde{g}_{12}-\sgn(g_1)\big)
\left(\bar{\psi}_2^2-\bar{\psi}_1^2\right)\bar{\psi}_1\bar{\psi}_2
+\tilde{U}_1\left(\bar{\psi}_2^4-\bar{\psi}_1^4\right)\right]\nonumber \\
&\pm 4\tilde{k}_0\bar{\psi}_1\bar{\psi}_2+\tilde{\Omega}_R\left(\bar{\psi}_2^2
-\bar{\psi}_1^2\right)=0.
\label{eq:GPsimple}
\end{align}
For the plane-wave phase, both $\bar{\psi}_1$ and $\bar{\psi}_2$ are assumed to
be constant, so that $\bar{\psi}_1^2+\bar{\psi}_2^2=\bar{n}$ and 
$\bar{\psi}_1^2-\bar{\psi}_2^2=\bar{n}s_z$. 
Inserting these into Eq.~(\ref{eq:GPsimple}) gives
\begin{align}
&\sqrt{1-s_z^2}\left[\mp 4\tilde{k}_0+s_z\frac{|g_1|\bar{n}}{E_R}
\Big(\tilde{g}_{12}-\sgn(g_1)\Big)\right]\nonumber \\
&+2s_z\left(\tilde{\Omega}_R+\tilde{U}_1\frac{|g_1|\bar{n}}{E_R}\right)=0.
\label{eq:GPsimple2}
\end{align}
When $\tilde{U}_1=0$ and $\tilde{\Omega}_R\approx 0$, this expression is 
approximately correct when $s_z\approx 1$, consistent with the variational
results in this regime. Recall that in the variational approach, the 
magnetization $s_z=\tilde{k}_0$ is constant [c.f.~Eq.~\eqref{eq:sz}], 
solely determined by $\tilde{\Omega}_R$. Unlike the variational result, 
however, it is immediately apparent from the second term in 
Eq.~(\ref{eq:GPsimple2}) that the magnetization must decrease monotonically as 
$\tilde{U}_1$ is increased. 

The magnetization $s_z$ obtained via numerical solution of 
Eq.~(\ref{eq:GPsimple2}) is shown as the red dotted curve in Fig.~\ref{fig:sz} 
for a condensate in the left well (i.e.\ choosing the lower sign) of the PWP
for $\tilde{U}_1\leq\tilde{U}_{1\rm c}$. Parameters are
$\tilde{U}_{1}=\tilde{U}_{\rm ss}=\tilde{U}_{\rm ds}$, $\sgn(g_1)=\tilde{g}_2
=|g_1|\bar{n}/E_R=\delta\tilde{U}=1$, $\tilde{g}_{12}=2$, and $\tilde{\Omega}_R=0.1$. 
As expected, the magnetization decreases monotonically with 
$\tilde{U}_1$ from its maximum at $\tilde{U}_1=0$. The difference between the
results of the two methods has its origins in the fact that the variational 
ansatz, Eq.~\eqref{eq:wavefunction-ansatz}, is a single-particle wavefunction 
which satisfies the GP equations in PWP only when all the two-body and 
cavity-mediated interactions are zero. In principle, the variational ansatz 
could be remedied by allowing both $k_0$ and $\theta_{\mathbf{k}_0}$ to be 
variational parameters~\cite{Li-2012}. The dependence of the solution of GP 
equations on the two-body and cavity-mediated interactions will be investigated
further in Sec.~\ref{sec:EX-PWP} in the calculation of the elementary 
excitations in the PWP. 


\subsubsection{Stripe phase}
\label{sec:SP}

The momentum dependence of the condensate in the SP is not as readily
apparent as it is for the PWP. It is therefore convenient to 
instead construct an effective low energy Hamiltonian by first mapping the 
complete Hamiltonian~\eqref{eq:totall-eff-H} into the lower band and then
deriving the low energy coupled GP equations~\cite{Lu-2013,Stanescu-2008}. 
This is reasonable because the occupation of the upper band 
$\epsilon_+(\mathbf{k})$ can be assumed to be small at low temperatures 
$k_{\rm B}T\ll\hbar\Omega_R$. Furthermore, only states in the vicinity of the 
two minima $\pm\tilde{\mathbf{k}}_0$ will be occupied. 

The field operators $\hat{\boldsymbol\Psi}(\mathbf r)$ can then be expanded in 
the lower band basis around the two minima (recall that 
$\boldsymbol\phi_-(\mathbf{k})$ is the \textit{two-component} spinor in the 
lower band):
\begin{align}
\hat{\boldsymbol\Psi}(\mathbf r)\simeq\sum_{\mathbf{q}<\mathbf{q}_c}
&\Big[e^{i(-\mathbf{k}_0+\mathbf{q})\cdot\mathbf{r}}
\boldsymbol\phi_-(-\mathbf{k}_0+\mathbf{q}) \hat{\varphi}_-(-\mathbf{k}_0+\mathbf{q})\nonumber\\
&+e^{i(\mathbf{k}_0+\mathbf{q})\cdot\mathbf{r}}
\boldsymbol\phi_-(\mathbf{k}_0+\mathbf{q}) \hat{\varphi}_-(\mathbf{k}_0+\mathbf{q})\Big],
\end{align} 
where the sum over $\mathbf{q}$ need only be taken up to some maximum 
$\mathbf{q}_c$. Approximating the spinor 
$\boldsymbol\phi_-(\pm\mathbf{k}_0+\mathbf{q})
\simeq\boldsymbol\phi_-(\pm\mathbf{k}_0)$ in the limit $\tilde{\Omega}_R\ll2$
and defining the new operators 
$\hat{\varphi}_{1'}(\mathbf{q})\equiv\hat{\varphi}_-(-\mathbf{k}_0+\mathbf{q})$
and $\hat{\varphi}_{2'}(\mathbf{q})\equiv\hat{\varphi}_
-(\mathbf{k}_0+\mathbf{q})$~\cite{Lu-2013}, the field operators read
\begin{align} \label{eq:low-energy-f-operator-mapping}
\hat{\boldsymbol\Psi}(\mathbf r)=
e^{-i\mathbf{k}_0\cdot\mathbf{r}} \boldsymbol\phi_-(-\mathbf{k}_0) \hat{\psi}_{1'}(\mathbf{r})
+e^{i\mathbf{k}_0\cdot\mathbf{r}} \boldsymbol\phi_-(\mathbf{k}_0) \hat{\psi}_{2'}(\mathbf{r}),
\end{align} 
where $\hat{\psi}_{\tau'}(\mathbf{r})=\sum_{\mathbf{q}} 
e^{i\mathbf{q}\cdot\mathbf{r}}\hat{\varphi}_{\tau'}(\mathbf{q})$.
In the small $\tilde{\Omega}_R$ limit and keeping terms only up to second order 
in $\tilde{\Omega}_R$ and noting that $k_0\simeq (1-\tilde{\Omega}_R^2/8)k_R$, 
the field operators can be further simplified to
\begin{align} \label{eq:low-energy-f-operator-mapping-simplified}
\begin{bmatrix}
\hat{\psi}_{1}(\mathbf{r}) \\
\hat{\psi}_{2}(\mathbf{r})
\end{bmatrix}
\simeq
\begin{bmatrix}
(1-\frac{\tilde{\Omega}_R^2}{32})e^{-ik_0z} &  \frac{\tilde{\Omega}_R}{4} e^{ik_0z}\\[.8mm]
-\frac{\tilde{\Omega}_R}{4} e^{-ik_0z} & -(1-\frac{\tilde{\Omega}_R^2}{32})e^{ik_0z}
\end{bmatrix}
\begin{bmatrix}
\hat{\psi}_{1'}(\mathbf{r}) \\
\hat{\psi}_{2'}(\mathbf{r})
\end{bmatrix}.
\end{align}
Note that the lab-frame pseudospin field operator $\hat{\tilde{\psi}}_{\tau}$ 
maps correctly to the corresponding dressed pseudospin field operator 
$\hat{\psi}_{\tau'}$ in the $\tilde{\Omega}_R\rightarrow0$ limit; recall that 
$\hat{\tilde{\boldsymbol\Psi}}(\mathbf r)
=\mathscr{U}_2^\dagger\hat{\boldsymbol\Psi}(\mathbf r)$. 
Substituting Eq.~\eqref{eq:low-energy-f-operator-mapping-simplified} back into 
the original Hamiltonian~\eqref{eq:totall-eff-H} and only keeping terms to 
second order in $\tilde{\Omega}_R$ yields the effective low-energy Hamiltonian:
\begin{align} \label{eq:low-energy-eff-H}
H_{\rm e}&=\int d^3r \left(\hat{\boldsymbol\Psi}'^\dagger H^{(1)}_{\rm e} \hat{\boldsymbol\Psi}'
+\frac {1}{2}g'_{1} \hat{n}_{1'}^2+\frac {1}{2}g'_{2} \hat{n}_{2'}^2+g'_{12} \hat{n}_{1'}\hat{n}_{2'} \right)\nonumber\\
&+\frac{1}{2} U'_{1} \hat{N}_{1'}^2+\frac{1}{2} U'_{2} \hat{N}_{2'}^2+ U'_{12} \hat{N}_{1'}\hat{N}_{2'},
\end{align}
where $ \hat{\boldsymbol\Psi}'(\mathbf{r})=(\hat{\psi}_{1'}(\mathbf{r}),\hat{\psi}_{2'}(\mathbf{r}))^\mathsf{T}$,
as before
$\hat{N}_{\tau'}=\int \hat{n}_{\tau'}(\mathbf r)d^3r
=\int \hat{\psi}_{\tau'}^\dagger(\mathbf r)\hat{\psi}_{\tau'}(\mathbf r) d^3r$ 
is the total atomic number operator for the dressed pseudospin $\tau'\in\{1',2'\}$,
and we have introduced the dressed interaction parameters
\begin{align} \label{eq:low-energy-H-coefficients}
g'_{\tau}&\equiv g_{\tau'\tau'}=g_{\tau}
-\frac{1}{8}(g_{\tau}-g_{12})\tilde{\Omega}_R^2,\nonumber\\
g'_{12}&\equiv g_{1'2'}=g_{12}
+\frac{1}{8}(g_{1}+g_{2})\tilde{\Omega}_R^2,\nonumber\\
U'_{\tau}&\equiv U_{\tau'\tau'}=2U_\tau-U_{\rm ds}\tilde{\Omega}_R
-\frac{1}{4}(U_\tau-U_{\rm ss})\tilde{\Omega}_R^2,\nonumber\\
U'_{12}&\equiv U_{1'2'}=-U_{\rm ds}\tilde{\Omega}_R
+\frac{1}{8}(U_1+U_2+2U_{\rm ss})\tilde{\Omega}_R^2,
\end{align}
with $\tau\in\{1,2\}$ and $\tau'\in\{1',2'\}$. 

The single-particle part of the effective low energy Hamiltonian 
$H^{(1)}_{\rm e}=(-\hbar^2/2m)[\nabla_{\!\bot\!}^2
+(1-\tilde{\Omega}_R^2/4)\partial_z^2]$
can be easily diagonalized~\cite{Lu-2013}, yielding the effective low energy dispersion 
$\epsilon_{\rm e}(\mathbf{k})/E_R=\tilde{k}_{\!\bot\!}^2
+(1-\tilde{\Omega}_R^2/4)\tilde{k}_z^2$. It is important to note that the 
lowest single-particle energy state for both dressed pseudospins is the 
$\mathbf{k}=0$ momentum state, not $\mathbf{k}=\pm \mathbf{k}_0$ as it was 
for the actual pseudospins. Then the effective low energy GP equations for the
SP can be obtained from $H_{\rm e}$, Eq.~\eqref{eq:low-energy-eff-H}:
\begin{align} \label{eq:ELE-GGPEs-SP}
\left[\left(\tilde{g}'_{1}+\tilde{U}'_1\right)|\psi_{1'}|^2
+\left(\tilde{g}'_{12}+\tilde{U}'_{12}\right)|\psi_{2'}|^2\right]\psi_{1'}
&=\bar{\mu} \psi_{1'},\nonumber\\
\left[\left(\tilde{g}'_{2}+\tilde{U}'_2\right)|\psi_{2'}|^2
+\left(\tilde{g}'_{12}+\tilde{U}'_{12}\right)|\psi_{1'}|^2\right]\psi_{2'}
&=\bar{\mu} \psi_{2'},
\end{align}
where the dressed pseudospin wavefunctions $\psi_{\tau'}$ are assumed to be 
homogeneous and unitless parameters have been introduced for convenience: 
$\tilde{g}'_{\tau}={g}'_{\tau}/|g_1|$, $\tilde{g}'_{12}={g}'_{12}/|g_1|$,
$\tilde{U}'_{\tau}=V{U}'_{\tau}/|g_1|$, and $\tilde{U}'_{12}=V{U}'_{12}/|g_1|$.
Here $\bar{\mu}=\mu/|g_1|$ which has units of inverse volume. 
These algebraic equations have the solution
\begin{align} \label{eq:SP-GGPEs-solution}
n_{1'}=\frac{2\tilde{U}_2+\tilde{g}'_2-\tilde{g}'_{12}
-\frac{1}{8}\left(\tilde{U}_1+3\tilde{U}_2\right)\tilde{\Omega}_R^2}
{\tilde{g}'_1+\tilde{g}'_2-2\tilde{g}'_{12}+2\left(\tilde{U}_1+\tilde{U}_2\right)
\left(1-\frac{1}{4}\tilde{\Omega}_R^2\right)}\bar{n},\nonumber\\
n_{2'}=\frac{2\tilde{U}_1+\tilde{g}'_1-\tilde{g}'_{12}
-\frac{1}{8}\left(3\tilde{U}_1+\tilde{U}_2\right)\tilde{\Omega}_R^2}
{\tilde{g}'_1+\tilde{g}'_2-2\tilde{g}'_{12}+2\left(\tilde{U}_1+\tilde{U}_2\right)
\left(1-\frac{1}{4}\tilde{\Omega}_R^2\right)}\bar{n}, 
\end{align}
where $n_{1'}+n_{2'}=\bar{n}$. Note that although the GP equations for the SP, 
Eq.~\eqref{eq:ELE-GGPEs-SP}, depend on the cavity parameters 
$\tilde{U}_{\rm ss}$ and $\tilde{U}_{\rm ds}$, these solutions do not; rather,
$\tilde{U}_1$ and $\tilde{U}_2$ are the only cavity interaction parameters that
affect $\psi_{\tau'}$, consistent with the variational approach of 
Sec.~\ref{sec:variational-approach}. 

The dressed magnetization $s'_z=(n_{1'}-n_{2'})/\bar{n}$ can easily be obtained
from Eq.~\eqref{eq:SP-GGPEs-solution}, and the actual magnetization 
$s_z=s'_z(1-\tilde{\Omega}_R^2/8)$ up to $\mathcal{O}(\tilde{\Omega}_R^3)$ is found using 
Eq.~\eqref{eq:low-energy-f-operator-mapping-simplified}: 
\begin{align} \label{eq:sz-SP}
s_z=\frac{\left[\tilde{g}'_2-\tilde{g}'_{1}
+2\delta\tilde{U}\left(1-\frac{1}{8}\tilde{\Omega}_R^2\right)\right]
\left(1-\frac{1}{8}\tilde{\Omega}_R^2\right)}
{\tilde{g}'_1+\tilde{g}'_2-2\tilde{g}'_{12}+2\left(\tilde{U}_1+\tilde{U}_2\right)
\left(1-\frac{1}{4}\tilde{\Omega}_R^2\right)}.
\end{align}
The SP magnetization $s_z$ is displayed as a function of 
$\tilde{U}_1$($\geqslant\tilde{U}_{1\rm c}$) in Fig.~\ref{fig:sz} with the 
red dotted curve for
$\sgn(g_1)=\tilde{g}_2=\delta\tilde{U}=1$, $\tilde{g}_{12}=2$, and 
$\tilde{\Omega}_R=0.1$. The behavior is indistinguishable from the 
magnetization obtained from the variational approach, Eq.~\eqref{eq:sz}. The
critical cavity interaction for the SP-PWP phase transition can be obtained 
from Eq.~\eqref{eq:SP-GGPEs-solution} by setting 
$n_{1'}=\bar{n}$ (or setting $s'_z=1$):
\begin{align} \label{eq:SP-left-PWP-tran-GGPE}
\tilde{U}_{1\rm c}^{\rm L}=\frac{1}{4(4-\tilde{\Omega}_R^2)}
\bigg\{&-\left[\tilde{g}_{12}-\sgn(g_1)-2\tilde{g}_2-\delta\tilde{U}\right]
\tilde{\Omega}_R^2
\nonumber\\
&+8\left[\tilde{g}_{12}-\sgn(g_1)\right]\bigg\}, 
\end{align}
for a phase transition from SP to a PWP at the left minimum. Instead setting 
$n_{1'}=0$ (or $s'_z=-1$) for a phase transition from SP to a PWP at the right 
minimum, one obtains
\begin{align} \label{eq:SP-right-PWP-tran-GGPE}
\tilde{U}_{1\rm c}^{\rm R}=\frac{1}{4(4-\tilde{\Omega}_R^2)}
\bigg\{&-\left[\tilde{g}_{12}-\sgn(g_1)-2\tilde{g}_2-3\delta\tilde{U}\right]
\tilde{\Omega}_R^2
\nonumber\\
&+8\left[\tilde{g}_{12}-\sgn(g_1)-2\delta\tilde{U}\right]\bigg\}.
\end{align}
Note that when $\sgn(g_1)=\tilde{g}_2$ and $\delta\tilde{U}=0$, the two 
critical cavity interactions $\tilde{U}_{1\rm c}^{\rm L}$ and 
$\tilde{U}_{1\rm c}^{\rm R}$ become equal to the value $\tilde{U}_{1\rm c}^0$ 
found using the variational approach, Eq.~\eqref{eq:SP-PWP-trans-VA}.

\subsection{Elementary Excitations: Bogoliubov theory}
\label{sec:elementary-excitation}

Thus far we have treated the bosons as classical fields, having replaced the 
field operators with their expectation values 
$\hat{\psi}_\tau \rightarrow \psi_\tau\equiv \langle \hat{\psi}_\tau \rangle$.
In this section, we consider the quantum fluctuations of the fields and obtain 
the elementary excitation spectrum using Bogoliubov theory. This is 
accomplished by writing the field operators as 
$\hat{\psi}_\tau=\psi_\tau+\delta\hat\psi_\tau$, where $\delta\hat\psi_\tau$ is 
the quantum fluctuation operator. These expressions are substituted into the
time-dependent GP equations and the resulting equations are linearized, i.e.\ 
terms are retained only up to first order in the fluctuations. One then obtains
a set of time-dependent coupled equations for $\delta\hat\psi_\tau$ which 
yields, after diagonalization, the elementary excitation spectrum.

\subsubsection{Plane wave phase}
\label{sec:EX-PWP}

Following the approach taken in Sec.~\ref{sec:PWP} for the PWP, it
is reasonable to define the bosonic field operator
\begin{align} \label{eq:psi-psibar}
\hat{\psi}_\tau(\mathbf{r},t)\equiv e^{\pm ik_0z}\left[\bar{\psi}_\tau
+\delta\hat{\psi}_\tau(\mathbf{r},t)\right],
\end{align} 
where $\bar{\psi}_\tau$ are the time-independent, homogeneous solutions of 
the coupled GP equations~\eqref{eq:GGPEs-PWP} in the PWP. 
To consider time-dependent fluctuations around the equilibrium solutions 
it is convenient to replace the chemical potential (which is the eigenvalue of 
the time-independent GP equations) by a time-dependent operator, 
$\mu\rightarrow\mu+i\hbar\partial_t$. The time-dependent fluctuations can then
be expressed using the usual Bogoliubov approach in terms of particle and hole 
excitations with amplitudes
$\bar{u}_{\tau,\mathbf{q}} e^{i(\mathbf{q}\cdot\mathbf{r}-\omega t)}$ and
$\bar{v}^*_{\tau,\mathbf{q}} e^{-i(\mathbf{q}\cdot\mathbf{r}-\omega t)}$,
respectively.

\begin{figure}[t]
\centering
\subfigure[]{
\includegraphics [width=0.225\textwidth]{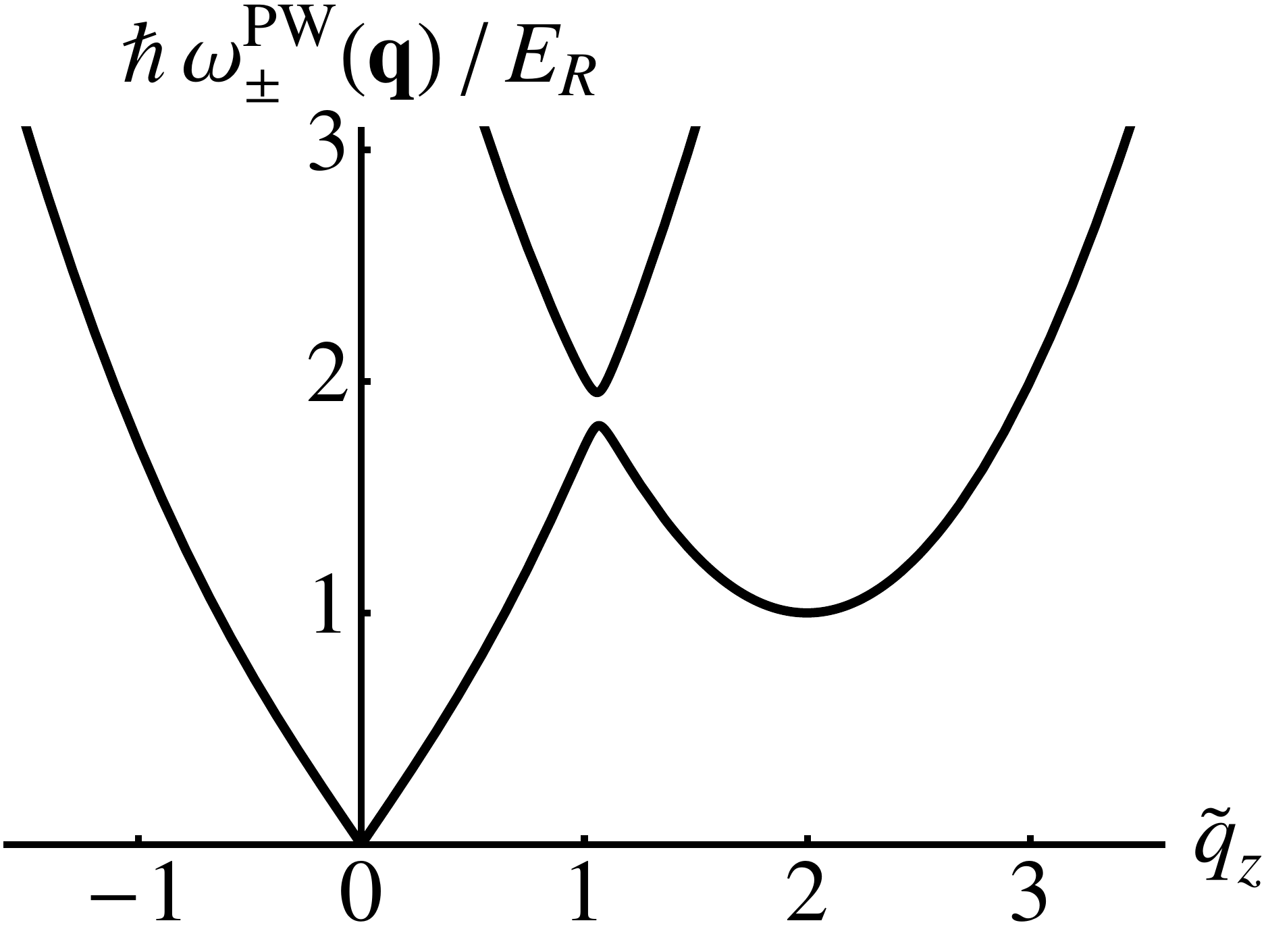}
\label{fig:elementary-excitation-PWP-no-cavity}}
\subfigure[]{
\includegraphics [width=0.225\textwidth]{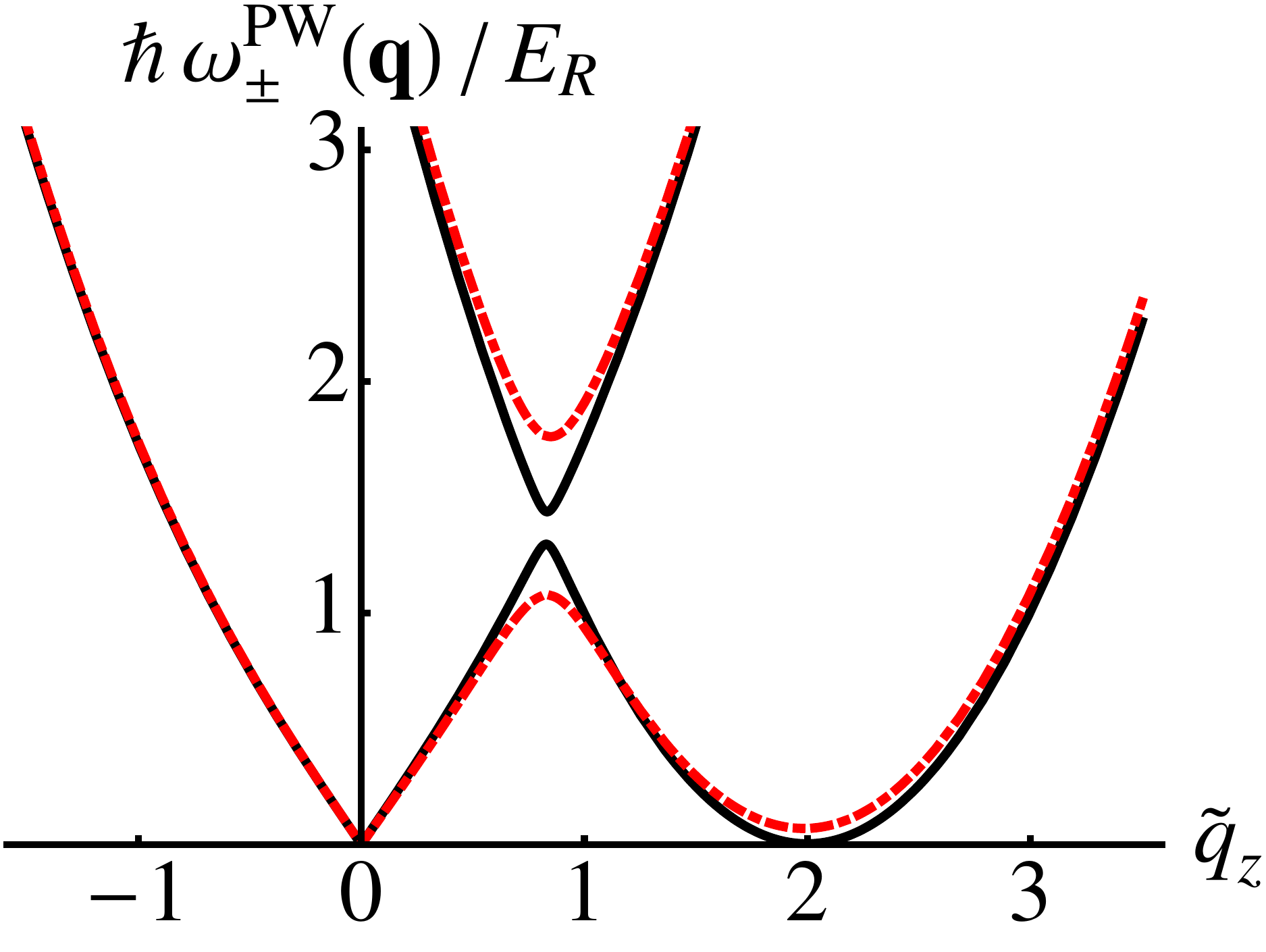}
\label{fig:elementary-excitation-PWP-with-cavity}}
\caption{(Color online) Elementary excitation spectrum in the PWP for
$\sgn(g_1)=\tilde{g}_2=|g_1|\bar{n}/E_R=1$, $\tilde{g}_{12}=2$, and 
$\tilde{\Omega}_R=0.1$. 
$(\tilde{U}_1,\tilde{U}_2,\tilde{U}_{\rm ss},\tilde{U}_{\rm ds})=(0,0,0,0)$ in (a),
and $(0.5,1.5,0,0)$ and $(0.5,1.5,0.5,0.5)$ in (b) for the black solid and 
red dashed-dotted curves, respectively.}
\label{fig:elementary-excitation-PWP}
\end{figure}

Consider the specific case of a condensate in the left minimum 
$-\mathbf{\tilde{k}}_0$ of the double-well single-particle 
dispersion relation; for condensation in the right well one need only replace
$\tilde{k}_0$ in what follows with $-\tilde{k}_0$. Substituting 
Eq.~\eqref{eq:psi-psibar} into the time-dependent GP equations and keeping only
linear terms in the fluctuations, one obtains the following non-Hermitian 
eigenvalue equation for each value of $\mathbf{q}$:

\begin{widetext}
\begin{align} \label{eq:Bogoliubov-H-PWP}
\begin{pmatrix}
M_{11} & g_1\bar\psi_1^2 & 
g_{12}\bar\psi_1\bar\psi^*_2+\hbar\Omega_{\rm eff}  & g_{12}\bar\psi_1\bar\psi_2 
\\[1mm]
-g_1\bar\psi_1^{*2} & -M_{22} & -g_{12}\bar\psi_1^*\bar\psi_2^* & 
-g_{12}\bar\psi_1^*\bar\psi_2-\hbar\Omega_{\rm eff}^* 
\\[1mm]
g_{12}\bar\psi_1^*\bar\psi_2+\hbar\Omega_{\rm eff}^* & 
g_{12}\bar\psi_1\bar\psi_2 & M_{33} & g_2\bar\psi_2^2 
\\[1mm]
-g_{12}\bar\psi_1^*\bar\psi_2^* & 
-g_{12}\bar\psi_1\bar\psi_2^*-\hbar\Omega_{\rm eff} & 
-g_2\bar\psi_2^{*2} & -M_{44}
\end{pmatrix}
\begin{pmatrix}
\bar{u}_{1,\mathbf{q}} \\[1mm]
\bar{v}_{1,\mathbf{q}} \\[1mm]
\bar{u}_{2,\mathbf{q}} \\[1mm]
\bar{v}_{2,\mathbf{q}}
\end{pmatrix}
=\hbar\omega(\mathbf{q})
\begin{pmatrix}
\bar{u}_{1,\mathbf{q}} \\[1mm]
\bar{v}_{1,\mathbf{q}} \\[1mm]
\bar{u}_{2,\mathbf{q}} \\[1mm]
\bar{v}_{2,\mathbf{q}}
\end{pmatrix},
\end{align} 
\end{widetext}
where 
\begin{align} \label{eq:Bogoliubov-matrix-element-PWP}
M_{11/22}&=E_R\left[\tilde{q}^2\mp2(\tilde{k}_0-1)\tilde{q}_z\right]+g_1|\bar\psi_1|^2-
\hbar\Omega_{\rm eff}\frac{\bar\psi_2}{\bar\psi_1},\nonumber\\
M_{33/44}&=E_R\left[\tilde{q}^2\mp2(\tilde{k}_0+1)\tilde{q}_z\right]+g_2|\bar\psi_2|^2-
\hbar\Omega_{\rm eff}^*\frac{\bar\psi_1}{\bar\psi_2},\nonumber\\
\hbar\Omega_{\rm eff}&=\hbar\Omega_R
+|g_1|\tilde{U}_{\rm ds}\bar{n}+2|g_1|\tilde{U}_{\rm ss} \bar{\psi}_1\bar{\psi}_2^*.
\end{align}
In deriving the Bogoliubov Hamiltonian \eqref{eq:Bogoliubov-H-PWP}, we made use
of the fact that $\hat{N}_{\tau}
=\int \hat{\psi}_{\tau}^\dagger(\mathbf{r},t)\hat{\psi}_{\tau}(\mathbf{r},t) 
d\mathbf{r}
=\int |\bar{\psi}_{\tau}|^2 d\mathbf{r}=V|\bar{\psi}_{\tau}|^2=N_\tau$,
because $\bar{\psi}_{\tau}$ is homogeneous by assumption and
$\int \delta\hat{\psi}_{\tau}(\mathbf{r},t) d\mathbf{r}=0$ because the 
spatial integral of either Bogoliubov amplitude $\bar{u}_{\tau,\mathbf{q}}
e^{i(\mathbf{q}\cdot\mathbf{r}-\omega t)}$ or $\bar{v}^*_{\tau,\mathbf{q}}
e^{-i(\mathbf{q}\cdot\mathbf{r}-\omega t)}$ is zero for any $\mathbf{q}\neq 0$.
A similar argument ensures that $\hat{S}_+=S_+$ and $\hat{S}_-=S_-$ as well. 
Note also that the chemical potential in Eq.~(\ref{eq:Bogoliubov-H-PWP}) has 
been eliminated using the coupled GP equations~(\ref{eq:GGPEs-PWP}).

Diagonalizing Eq.~\eqref{eq:Bogoliubov-H-PWP} yields the spectrum
$\omega^{\rm PW}_{\pm}(\mathbf{q})$ of collective excitations. The results are 
shown in Fig.~\ref{fig:elementary-excitation-PWP-no-cavity} for the parameters 
$\sgn(g_1)=\tilde{g}_2=|g_1|\bar{n}/E_R=1$, $\tilde{g}_{12}=2$, and 
$\tilde{\Omega}_R=0.1$, when all the cavity-mediated interaction terms are zero
($\tilde{U}_1=\tilde{U}_2=\tilde{U}_{\rm ss}=\tilde{U}_{\rm ds}=0$), i.e.\ the 
system is deep in the PWP. The lower curve exhibits the usual superfluid 
sound-like linear dispersion around the origin $\tilde{q}_z\equiv q_z/k_R=0$ 
(around the left minimum of the single-particle energy dispersion where all the 
atoms are condensed) and a roton-type minimum around $\tilde{q}_z\simeq 2$. As 
the cavity interactions are increased, the energy of the roton minimum lowers.
For parameters $\tilde{U}_1=0.5$, $\delta\tilde{U}=1.5$, 
$\tilde{U}_{\rm ss}=\tilde{U}_{\rm ds}=0$, and the other parameters same as in
Fig.~\ref{fig:elementary-excitation-PWP-no-cavity}, this minimum coincides with 
zero energy (i.e.~the excitation energy at the origin $\tilde{q}_z=0$); see the 
black solid curve in Fig.~\ref{fig:elementary-excitation-PWP-with-cavity}. The 
red dashed-dotted curve represents the elementary excitation spectrum for the 
same values of $\tilde{U}_1$ and $\delta\tilde{U}$ but for 
$\tilde{U}_{\rm ss}=\tilde{U}_{\rm ds}=0.5$. In this case, 
$\hbar\Omega_{\rm eff}/E_R$ [cf.\ Eq.~(\ref{eq:Bogoliubov-matrix-element-PWP})]
is somewhat bigger than the bare $\tilde\Omega_R=0.1$ for the black solid curve,
so the roton minimum lies somewhat above that of the black solid curve around $\tilde{q}_z\simeq2$.

The energy of the roton minimum near $q_z\simeq 2k_R$ can be reduced
below zero by further increasing the cavity interaction strength 
$\tilde{U}_1$. This signals a dynamic instability toward the
formation of the SP; recall from Eq.~(\ref{eq:tot-density}) that the 
density modulation in the SP has wave vector $2k_0\simeq2k_R$ 
for $\tilde{\Omega}_R\rightarrow0$. The critical cavity interactions
for the black solid and the red dashed-dotted excitation spectra 
in Fig.~\ref{fig:elementary-excitation-PWP-with-cavity} are 
$\tilde{U}_{1 \rm c}\simeq0.5$ and $0.53$, respectively,  
and these are in good agreement with that of the variational 
approach, where Eq.~(\ref{eq:Ucbetter}) predicts a phase transition between 
the PWP and the SP at the critical value $\tilde{U}_{1\rm c}\simeq0.5$ for 
the parameters $\sgn(g_1)=\tilde{g}_2=\delta\tilde{U}=1$, $\tilde{g}_{12}=2$, 
and $\tilde{\Omega}_R=0.1$ (cf.~also Fig.~\ref{fig:sz}).

If one hypothetically sets $\tilde{U}_{\rm ss}=\tilde{U}_{\rm ds}=0$ in the 
PWP, then the critical cavity interaction $\tilde{U}_{1\rm c}$ obtained 
from the analysis of the elementary excitations and the variational 
method would match exactly with each other for any range of parameters.
Nevertheless, they begin to deviate from one another as
$\tilde{U}_{\rm ss}$ and $\tilde{U}_{\rm ds}$ become larger and larger, 
because Eq.~(\ref{eq:Ucbetter}) is independent of these cavity
interaction parameters while both the coupled GP
equations and the Bogoliubov Hamiltonian depend 
explicitly on them (the latter through $\hbar\Omega_{\rm eff}$). 
That said, we have compared the critical 
phase transition point $\tilde{U}_{1\rm c}$ obtained from both the 
variational approach and the elementary excitation spectrum in the PWP and have
found that when $\tilde{U}_1=\tilde{U}_{\rm ss}=\tilde{U}_{\rm ds}$ 
they agree with one another within a $\sim 8\%$ error for $\tilde{g}_{12}$ 
in the range of $\sim0-8$, assuming 
$\sgn(g_1)=\tilde{g}_2=|g_1|\bar{n}/E_R=\delta\tilde{U}=1$ and 
$\tilde{\Omega}_R=0.1$.

\subsubsection{Stripe phase}
\label{sec:EX-SP}

The derivation of the Bogoliubov excitation spectrum begins with the 
corresponding time-dependent, effective low energy GP equations in the SP
[c.f.\ Eq.~(\ref{eq:ELE-GGPEs-SP})]:
\begin{widetext}
\begin{align} \label{eq:TD-ELE-GGPEs-SP}
i\hbar\frac{\partial}{\partial t}\hat{\psi}_{1'}=
&\Big(H^{(1)}_{\rm e}+
g'_{1}|\hat{\psi}_{1'}|^2+g'_{12}|\hat{\psi}_{2'}|^2+U'_1\hat{N}_{1'}
+U'_{12}\hat{N}_{2'}-\mu\Big)\hat{\psi}_{1'},\nonumber\\
i\hbar\frac{\partial}{\partial t}\hat{\psi}_{2'}=
&\Big(H^{(1)}_{\rm e}+
g'_{2}|\hat{\psi}_{2'}|^2+g'_{12}|\hat{\psi}_{1'}|^2+U'_2\hat{N}_{2'}
+U'_{12}\hat{N}_{1'}-\mu\Big)\hat{\psi}_{2'}.
\end{align}
As in the PWP case, the low energy field operators are replaced with 
$\hat{\psi}_{\tau'}(\mathbf{r},t)=\psi_{\tau'}
+\delta\hat{\psi}_{\tau'}(\mathbf{r},t)$ in these equations. Here 
$\psi_{\tau'}$ are the time-independent, homogeneous solutions of the effective 
low energy GP equations in the SP, Eq.~\eqref{eq:SP-GGPEs-solution},  and 
$\delta\hat{\psi}_{\tau'}(\mathbf{r},t)$ are the quantum fluctuations.
Linearizing Eq.~\eqref{eq:TD-ELE-GGPEs-SP} yields the Bogoliubov Hamiltonian:
\begin{align} \label{eq:Bogoliubov-H-PWP2}
\begin{pmatrix}
\epsilon_{\rm e}(\mathbf{q})+g'_1|\psi_{1'}|^2 & g'_1\psi_{1'}^2 & 
g'_{12}\psi_{1'}\psi^*_{2'} & g'_{12}\psi_{1'}\psi_{2'} 
\\[1mm]
-g'_1\psi_{1'}^{*2} & -\epsilon_{\rm e}(\mathbf{q})-g'_1|\psi_{1'}|^2 & 
-g'_{12}\psi_{1'}^*\psi_{2'}^* & -g'_{12}\psi_{1'}^*\psi_{2'} 
\\[1mm]
g'_{12}\psi_{1'}^*\psi_{2'} & g'_{12}\psi_{1'}\psi_{2'} 
& \epsilon_{\rm e}(\mathbf{q})+g'_2|\psi_{2'}|^2 & g'_2\psi_{2'}^2 
\\[1mm]
-g'_{12}\psi_{1'}^*\psi_{2'}^* & -g'_{12}\psi_{1'}\psi_{2'}^* & 
-g'_2\psi_{2'}^{*2} & -\epsilon_{\rm e}(\mathbf{q})-g'_2|\psi_{2'}|^2
\end{pmatrix}
\begin{pmatrix}
u_{1',\mathbf{q}} \\[1mm]
v_{1',\mathbf{q}} \\[1mm]
u_{2',\mathbf{q}} \\[1mm]
v_{2',\mathbf{q}}
\end{pmatrix}
=\hbar\omega(\mathbf{q})
\begin{pmatrix}
u_{1',\mathbf{q}} \\[1mm]
v_{1',\mathbf{q}} \\[1mm]
u_{2',\mathbf{q}} \\[1mm]
v_{2',\mathbf{q}}
\end{pmatrix},
\end{align} 
\end{widetext}
which can be diagonalized to give the spectrum of the elementary excitations:
\begin{align} \label{eq:excitation-spectrum-SP}
\hbar\omega_{\pm}^{\rm SP}(\mathbf{q})=\sqrt{\epsilon_{\rm e}^2(\mathbf{q})
+\epsilon_{\rm e}(\mathbf{q})\left(D_1\pm\sqrt{D_1^2-4D_2}\right)},
\end{align}
with
\begin{align}
D_1&=g'_1n_{1'}+g'_2n_{2'},\nonumber\\
D_2&=(g'_1g'_2-g'^2_{12})n_{1'}n_{2'}.
\end{align}
We have again used the fact that $\hat{N}_{\tau'}=N_{\tau'}$. 

Surprisingly,
the Bogoliubov Hamiltonian in the SP does not depend explicitly on the cavity 
parameters and the form of the excitation spectrum coincides with the 
quasiparticle spectrum of a Raman-induced stripe phase BEC~\cite{Lu-2013}. That 
said, the excitation spectrum implicitly depends on the cavity parameters
$\tilde{U}'_{\tau}$ through $n_{\tau'}$, as can be seen in
Eq.~\eqref{eq:SP-GGPEs-solution}. Both $\omega_{\pm}^{\rm SP}(\mathbf{q})$ are 
gapless and exhibit linear dispersion at long wavelengths, the characteristic 
of superfluidity in this phase; the slope of the dispersion relation at long
wavelength corresponds to the speed of sound in the medium. In the transverse
direction, one obtains
\begin{align} 
v_{\!\bot\!}^{(\pm)}=\frac{d\omega_{\pm}^{\rm SP}(\mathbf{q})}{dq_{\bot\!}}\Big|_{\mathbf{q}\rightarrow0}
=\frac{1}{\sqrt{2m}}\sqrt{D_1\pm\sqrt{D_1^2-4D_2}},
\end{align}
and the speed of sound in the $z$ (SO-coupling) direction is nearly the same
for small $\tilde{\Omega}$, 
$v_{z}^{(\pm)}=v^{(\pm)}_{\!\bot\!}\sqrt{1-\tilde{\Omega}_R^2/4}$.

\begin{figure}[t]
\centering
\includegraphics [width=0.45\textwidth]{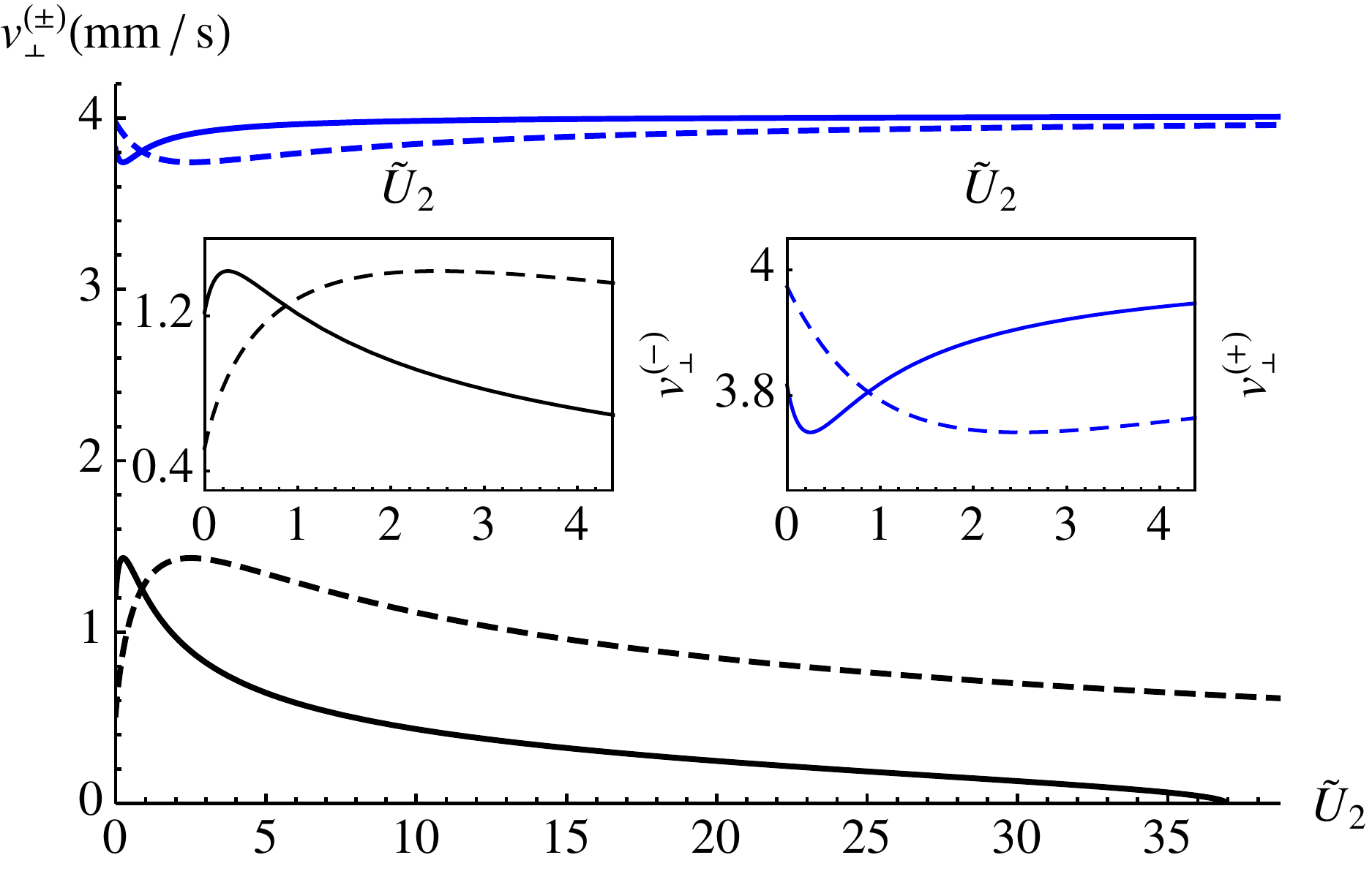}
\caption{(Color online) The speed of sound in the transverse direction 
$v^{(\pm)}_{\!\bot\!}$ is shown as a function of $\tilde{U}_2$ for 
$\tilde{U}_1=1/4$ (solid curves) and $\tilde{U}_1=5/2$ (dashed curves). For all 
curves: $\tilde{\Omega}_R=0.4$, $\sgn(g_1)=\tilde{g}_2=1$, 
$\tilde{g}_{12}=0.7$, $g_1\bar{n}/E_R=1$, and $m$ is the mass of $^{87}$Rb 
atom. The insets show the results closer to the origin.}
\label{fig:sound_speed_SP}
\end{figure}

Figure~\ref{fig:sound_speed_SP} depicts $v^{(\pm)}_{\!\bot\!}$ as a function of 
$\tilde{U}_2$ for $\tilde{U}_1=1/4$ (solid curves) and $\tilde{U}_1=5/2$ 
(dashed curves), with the other parameters fixed to $\sgn(g_1)=\tilde{g}_2=1$, 
$\tilde{g}_{12}=0.7$, $\tilde{\Omega}_R=0.4$, and $g_1\bar{n}/E_R=1$. The mass 
is assumed to be that of $^{87}$Rb. As $\tilde{U}_2$ is increased above zero, 
the speed of sound in the positive branch $v^{(+)}_{\!\bot\!}$ (the blue 
curves) first decreases quickly and reaches a minimum around 
$\delta\tilde{U}=\tilde{U}_2-\tilde{U}_1\sim 0$ for both curves, and then 
gradually approaches its asymptotic value. The speed of sound in the negative 
branch $v^{(-)}_{\!\bot\!}$ (black curves) has the opposite behavior, 
first increasing sharply to a maximum again near $\delta\tilde{U}\sim 0$ for 
both curves, before asymptotically approaching zero. The insets show the 
behaviour of $v^{(\pm)}_{\!\bot\!}$ close to the origin. The asymptotic 
behaviour of the speed of sound can be understood by noting that for large 
positive $\tilde{U}_2\gg\tilde{U}_1$, $n_{1'}$ approaches $\bar{n}$ and
$n_{2'}$ approaches zero [c.f.\ Eqs.~(\ref{eq:SP-GGPEs-solution})].
As a consequence $D_2\rightarrow0$ and 
$v^{(-)}_{\!\bot\!}\rightarrow0$ while 
$v^{(+)}_{\!\bot\!}\rightarrow\sqrt{g_{1}'\bar{n}/m}$. For the solid curves 
(where $\tilde{U}_1=1/4$), the speed of sound in the negative branch 
$v^{(-)}_{\!\bot\!}$ becomes zero at $\tilde{U}_2\simeq 37$, consistent with
the value at which the dressed magnetization $s'_z$ becomes 
unity for this choice of parameters. This signifies an instability toward the 
formation of a different phase. 

The condition that the speed of sound must be non-negative imposes the 
constraint $D_2\geqslant0$. This condition marks the onset of a phase 
transition at the critical point 
$\tilde{g}'^{\rm (c)}_{12}=\sqrt{\tilde{g}_1'\tilde{g}_2'}$, which does not 
depend on any cavity-mediated interaction parameters and is solely determined 
by the two-body interactions and $\tilde{\Omega}_R$. This critical point is not 
consistent with the previous results obtained from the variational approach, 
the effective low-energy GP equations in the SP, or the elementary excitations 
in the PW which all consistently predict a critical point for the PWP-SP phase
transition that depends on the cavity-mediated interaction parameters. To
verify that there was not an error in the calculations, the elementary 
excitations were computed directly in momentum space by Fourier transferring 
the effective low-energy Hamiltonian~\eqref{eq:low-energy-eff-H}, and treating 
the fluctuations around the condensate $\varphi_{\tau'}(\mathbf{q}=0)$ to 
second order in $\hat{\varphi}_{\tau'}(\mathbf{q})$ for small momenta 
$\mathbf{q}$. The results were identical with the real-space analysis, 
Eq.~\eqref{eq:excitation-spectrum-SP}. Interestingly, the critical 
inter-species interaction $\tilde{g}'^{\rm (c)}_{12}$ above defines a phase 
boundary between the stripe phase and a phase-separated state in Raman-induced 
spin-orbit coupled BECs~\cite{Lu-2013}. It is therefore conceivable that there 
is another phase between the SP and the PWP induced by the cavity interactions, 
whose signature is the observed inconsistency in the critical point.

\section{Discussion and Conclusions}
\label{sec:conclusions}

In this work we have shown that cavity-mediated long-ranged interactions
between atoms can profoundly alter the nature of the ground state and the 
elementary excitations of a cavity-induced spin-orbit-coupled two-component 
BEC, for ring-type cavities in the weak-coupling regime. Specifically, 
experimentally tunable
cavity-mediated interactions compete with the standard two-body interactions 
to yield both plane-wave and stripe phase ground states. Indeed, positive 
long-range cavity interactions can stabilize fully attractive BECs (condensates
where intra-species collisional interactions are negative, independent of the 
sign of the inter-species interaction) against collapse in the stripe phase. 
The collective excitations of the plane-wave phase ground states are found to
have a distinctive roton-type excitation spectrum reminiscent of that of 
superfluid $^4$He, which can be used as a signature of the phase. The stripe 
phase has a standard linear dispersion relation; the associated speed of sound
is found to go negative at a critical value of the cavity interaction strength,
signalling an instability toward another (likely phase-separated) phase. The
results suggest that cavity QED, even in the weak-coupling regime, can yield
interesting new physics for spin-orbit coupled BECs.

The results raise interesting avenues for future investigations. This work
assumed a fictional experimental configuration where the momentum is a good 
quantum number in the direction of the applied spin-orbit interactions. In
reality the condensate would be confined in this direction, and even a weak
harmonic potential could change the physics. While the stripe phase would 
likely remain robust, as it is essentially a weak standing wave superimposed on
the background condensate density profile, the plane-wave phase has no analog
in a confined geometry. Another loose end is the nature of the phase hinted at
in the limit of a large difference $\delta\tilde{U}$ between the 
cavity-mediated interactions between the two kinds of spin components 
$\tilde{U}_1$ and $\tilde{U}_2$. For large $\delta\tilde{U}$, the sound 
velocity in the stripe phase was found to go negative, a signature of the 
dynamic instability of the phase. While other work suggests that this signals
a 

However, a few intriguing issues and questions remain unclear and deserve further investigations. 
These include the inconsistency in the critical phase transition point, 
how the combined SO coupling effect, the two-body interactions, 
and the cavity-mediated long-ranged 
interactions change the superfluid--Mott-insulator phase transition 
as well as the nature of magnetic orders in the Mott-insulating regime
when an optical lattice imposed inside the cavity. Furthermore, whether it is possible 
to have a superfluid--Mott-insulator-like phase transition with solely the cavity-mediated 
long-range interactions, whether there is more interesting physics in strong-coupling
regime, and how the cavity fields are affected by the atoms.    
Some of these questions are the subject of our current works
with some promising preliminary results and will be published elsewhere.

\begin{acknowledgements}
The authors are grateful to Paul Barclay and Christoph Simon  for constructive 
criticisms. We also thank Han Pu and Lin Dong for stimulating correspondence.
This work was supported by the Natural Sciences and Engineering Research 
Council of Canada and Alberta Innovates-Technology Futures.
\end{acknowledgements}

\appendix
\begin{widetext}

\section{Adiabatic Elimination of the Atomic Excited State}
\label{app:atomic-adiabatic-elimination}

We first express the single-particle Hamiltonian density $\mathcal{H}^{(1)}$, Eq.~\eqref{total-single-atom-H}, 
in the rotating frame of pump lasers \cite{Maschler-2008} by applying the unitary transformation
\begin{align*} 
\mathscr{U}_1=
\exp \left\{ i \left[\left(\hat{A}_1^\dagger \hat{A}_1-\sigma_{aa}\right)\omega_{\rm p1}+
\left(\hat{A}_2^\dagger \hat{A}_2-\sigma_{bb}\right)\omega_{\rm p2}\right]t\right\},
\end{align*}
to obtain
\begin{align} \label{eq:SM-1st-UT-applied-H}
\mathcal{H}'^{(1)}&=\left[\frac{\mathbf{p}^2}{2m}+V_{\rm ext}(\mathbf{r})\right]I_{3\times3}
+\frac{\hbar\delta'}{2} \left(\sigma_{aa}- \sigma_{bb}\right)
-\frac{\hbar}{2}\left(\Delta_{\rm a1}+\Delta_{\rm a2}\right)\sigma_{ee}
+\hbar \left[\left(\mathscr{G}_{ae}e^{ik_Rz}\hat{A}_1\sigma_{ea}+
\mathscr{G}_{be}e^{-ik_Rz}\hat{A}_2\sigma_{eb} \right)+ \text{H.c.}\right]\nonumber\\
&-\hbar\big(\Delta_{\rm c1}\hat{A}_1^\dagger \hat{A}_1+\Delta_{\rm c2}\hat{A}_2^\dagger \hat{A}_2\big)
+i\hbar\left[\big(\eta_1 \hat{A}_1^\dagger + \eta_2 \hat{A}_2^\dagger\big)- \text{H.c.} \right],
\end{align}
where we have defined the atomic and the two-photon (or relative-atomic) detunings
\begin{subequations} \label{eq:SM-detunings}
\begin{align}
\Delta_{\rm a1}=\omega_{\rm p1}-\frac{1}{\hbar}(\varepsilon_e-\varepsilon_a),\qquad
\Delta_{\rm a2}=\omega_{\rm p2}-\frac{1}{\hbar}(\varepsilon_e-\varepsilon_b),\qquad
\delta'=(\omega_{\rm p1}-\omega_{\rm p2})-\frac{1}{\hbar}(\varepsilon_b-\varepsilon_a)
=\Delta_{\rm a1}-\Delta_{\rm a2},
\end{align}
{and cavity detunings}
\begin{align}
\Delta_{{\rm c} j}=\omega_{{\rm p}j}-\omega_j, \qquad j =1,2,
\end{align}
\end{subequations}
with respect to the pump lasers. 
Let us now assume that the detunings 
$\Delta_1=\omega_1-\varepsilon_{ea}/\hbar=-\Delta_{\rm c1}+\Delta_{\rm a1}$ and  
$\Delta_2=\omega_2-\varepsilon_{eb}/\hbar=-\Delta_{\rm c2}+\Delta_{\rm a2}$
are large compared to 
$\varepsilon_{ba}/\hbar=(\varepsilon_b-\varepsilon_a)/\hbar$ so that we can adiabatically eliminate 
the dynamic of the atomic excited state $\ket{e}$ from the Hamiltonian 
\eqref{eq:SM-1st-UT-applied-H} and obtain an effective Hamiltonian for
the ground pseudospins $\{1,2\}\equiv\{b,a\}$. 
Following the standard adiabatic elimination procedure 
\cite{Gerry-1990,Mivehvar-2014}, we first find the Heisenberg equations of motion
$i\hbar\dot{\sigma}_{e\tau}=[\sigma_{e\tau},\mathcal{H}'^{(1)}]$ for $\dot{\sigma}_{ea}$ and
$\dot{\sigma}_{eb}$, and then (after transferring to slowly rotating variables) 
set them equal to zero to find the steady-state solutions
$\sigma_{ea}^{\rm (ss)}$ and $\sigma_{eb}^{\rm (ss)}$. After substituting these steady-state
solutions back in $\mathcal{H}'^{(1)}$ \eqref{eq:SM-1st-UT-applied-H} and dropping 
terms diagonal in $\sigma_{ee}$, we arrive at the single-particle Hamiltonian density
for pseudospins
\begin{align} \label{eq:SM-adiabatic-eliminated-H}
\mathcal{H}_{\rm SO}'^{(1)}&=\left[\frac{\mathbf{p}^2}{2m}+V_{\rm ext}(\mathbf{r})\right]\mathbb{I}
+\hat{\varepsilon}_{1}\sigma_{11}
+\hat{\varepsilon}_{2}\sigma_{22}
+\hbar\Omega'_R\Big(e^{2ik_Rz} \hat{A}_2^\dagger \hat{A}_1 \sigma_{12}
+e^{-2ik_Rz} \hat{A}_1^\dagger \hat{A}_2 \sigma_{21}\Big)+H'_{\rm cav},
\end{align}
where
\begin{align*}
H'_{\rm cav}=
-\hbar\big(\Delta_{\rm c1}\hat{A}_1^\dagger \hat{A}_1+\Delta_{\rm c2}\hat{A}_2^\dagger \hat{A}_2\big)
+i\hbar\left[\left(\eta_1 \hat{A}_1^\dagger + \eta_2 \hat{A}_2^\dagger\right)- \text{H.c.}\right],
\end{align*}
and
\begin{align} \label{eq:SM-1st-adiabatic-pseudospin-energies}
\hat{\varepsilon}_{1}=-\frac{\hbar \delta'}{2}+
\frac{2\hbar \mathscr{G}_{be}^2}{\Delta_2}(\hat{A}_2^\dagger \hat{A}_2+\frac{1}{2}),\qquad
\hat{\varepsilon}_{2}=\frac{\hbar \delta'}{2}+
\frac{2\hbar \mathscr{G}_{ae}^2}{\Delta_1}(\hat{A}_1^\dagger \hat{A}_1+\frac{1}{2}).
\end{align}
Here, $\Omega'_R=\frac{\Delta_1+\Delta_2}{\Delta_1\Delta_2}\mathscr{G}_{ae}\mathscr{G}_{be}$ 
is the two-photon Rabi frequency and $\mathbb{I}\equiv I_{2\times2}$ is the identity matrix in the pseudospin space.
Note the hat on $\hat{\varepsilon}_{\tau}$, implying that it depends on the cavity field operators.
After transferring to the co-moving frame of the cavity modes by applying the unitary transformation
$\mathscr{U}_2=e^{-ik_R(\sigma_{11} - \sigma_{22})z}$ 
to the Hamiltonian density \eqref{eq:SM-adiabatic-eliminated-H},
we obtain the SO-coupled single-particle Hamiltonian density  
\begin{align} \label{eq:SM-2nd-UT-applied-H}
\mathcal{H}_{\rm SO}''^{(1)}&=
\frac{1}{2m}\left\{p_{\!\bot\!}^2\mathbb{I}+\big[p_z\mathbb{I}-\hbar k_R (\sigma_{22} - \sigma_{11}) \big]^2 \right\}
+V_{\rm ext}(\mathbf{r})\mathbb{I}
+\sum_{\tau=1,2}\hat{\varepsilon}_{\tau}\sigma_{\tau\tau}
+\hbar\Omega'_R\Big(\hat{A}_2^\dagger \hat{A}_1 \sigma_{12}
+\hat{A}_1^\dagger \hat{A}_2 \sigma_{21}\Big)
+H'_{\rm cav}.
\end{align}
One can identify $\hbar k_R (\sigma_{22} - \sigma_{11})$ 
with $eA^*_z/c$ as in the minimal coupling Hamiltonian, that is, 
$eA^*_z/c\equiv \hbar k_R (\sigma_{22} - \sigma_{11})=-\hbar k_R\sigma_z$,
 where $\sigma_z=\sigma_{11}-\sigma_{22}$ is the third Pauli matrix. 
Nonetheless, we emphasis that here $A^*_z$ is a matrix acting in the internal pseudospin states, 
in contrast to the ordinary vector potential whose components are scaler fields.
Then the single-particle Hamiltonian reads
\begin{align} \label{eq:SM-2nd-UT-applied-H-integrated}
H_{\rm SO}''^{(1)}=\frac{1}{2m}\int \hat{\boldsymbol\Psi}^\dagger 
\left[p_{\!\bot\!}^2\mathbb{I}+(p_z\mathbb{I}+\hbar k_R \sigma_z )^2
+V_{\rm ext}(\mathbf{r})\mathbb{I}\right] \hat{\boldsymbol\Psi} d^3r
+\sum_{\tau=1,2}\hat{\varepsilon}_{\tau}\hat{N}_{\tau}
+\hbar\Omega'_R\Big(\hat{A}_2^\dagger \hat{A}_1 \hat{S}_+
+\hat{A}_1^\dagger \hat{A}_2 \hat{S}_-\Big)
+H'_{\rm cav},
\end{align}
where $\hat{\boldsymbol\Psi}(\mathbf r)=(\hat{\psi}_1(\mathbf r),\hat{\psi}_2(\mathbf r))^\mathsf{T}$ 
are the bosonic field operators, 
$\hat{N}_{\tau}=\int \hat{\psi}_{\tau}^\dagger(\mathbf r)\hat{\psi}_\tau(\mathbf r) d^3r$ 
is the total atomic number operator for pseudospin $\tau$,
$\hat{N}=\hat{N}_{1}+\hat{N}_{2}$ is the total atomic number operator, and 
$\hat{S}_+=\hat{S}_-^\dagger=\int \hat{\psi}_1^\dagger(\mathbf r)\hat{\psi}_2(\mathbf r) d^3r$
are the collective pseudospin raising and lowering operators. 

\section{Adiabatic Elimination of the Cavity Fields}
\label{app:cavity-adiabatic-elimination}

By noting that the cavity field operator commutes with the atomic interaction Hamiltonian 
$[\hat{A},H_{\rm int}]=0$, then the Heisenberg equations of motion of the cavity field operators 
are determined by the single-particle Hamiltonian $H_{\rm SO}''^{(1)}$, 
Eq.~\eqref{eq:SM-2nd-UT-applied-H-integrated}: 
$\partial_t{\hat{A}}_j=-i[\hat{A}_j,H_{\rm SO}''^{(1)}]/\hbar-\kappa\hat{A}_j$,
where the cavity-mode decay $-\kappa\hat{A}_j$ is included phenomenologically.
They can be recast in the matrix form,
\begin{align} \label{eq:SM-EOM-cavity-fields}
\frac{d}{dt}
\begin{pmatrix}
\hat{A}_1 \\
\hat{A}_2
\end{pmatrix}
=i
\begin{pmatrix}
\hat{\alpha}_{11} & -\hat{\alpha}_{12} \\
-\hat{\alpha}_{21} & \hat{\alpha}_{22}
\end{pmatrix}
\begin{pmatrix}
\hat{A}_1 \\
\hat{A}_2
\end{pmatrix}
+
\begin{pmatrix}
\eta_1 \\
\eta_2
\end{pmatrix},
\end{align}
where the elements of the "operator" matrix $\hat{\boldsymbol\alpha}$ are given by
\begin{align} \label{eq:SM-alpha-matrix}
\hat{\alpha}_{11}=(\Delta_{\rm c1}+i\kappa)-\frac{2\mathscr{G}_{ae}^2}{\Delta_1}\hat{N}_{2},\qquad
\hat{\alpha}_{22}=(\Delta_{\rm c2}+i\kappa)-\frac{2\mathscr{G}_{be}^2}{\Delta_2}\hat{N}_{1},\qquad
\hat{\alpha}_{12}=\hat{\alpha}_{21}^\dagger=\Omega'_R \hat{S}_-.
\end{align}
If the cavity decay rate $\kappa$ is large, then the cavity fields reach steady states very quickly. 
By setting $\partial_t{\hat{A}}_1=\partial_t{\hat{A}}_2=0$ in 
Eq.~\eqref{eq:SM-EOM-cavity-fields}, one can simultaneously solve the two equations of motion 
to obtain formal expressions for the steady-state field amplitudes $\hat{A}_{{\rm ss}j}$. 
However, one should take special care in solving these equations since the cavity fields and 
atomic fields commute with one another and this can give rise to ambiguities in solving these 
equations. In order to avoid such ambiguities, we symmetrize the equations of motion 
and exercise symmetrization procedure in all results following from the equations of motion. 
Thus, after setting $\partial_t{\hat{A}}_1=\partial_t{\hat{A}}_2=0$ in 
Eq.~\eqref{eq:SM-EOM-cavity-fields},
we re-express equations of motion as
\begin{align} \label{eq:SM-EOM-cavity-fields-symmetrized}
\frac{i}{2}\left(\hat{\alpha}_{11}\hat{A}_{{\rm ss}1}+\hat{A}_{{\rm ss}1}\hat{\alpha}_{11}\right)
-\frac{i}{2}\left(\hat{\alpha}_{12}\hat{A}_{{\rm ss}2}+\hat{A}_{{\rm ss}2}\hat{\alpha}_{12} \right)+\eta_1
=
\frac{i}{2}\left(\hat{\alpha}_{22}\hat{A}_{{\rm ss}2}+\hat{A}_{{\rm ss}2}\hat{\alpha}_{22}\right)
-\frac{i}{2}\left(\hat{\alpha}_{21}\hat{A}_{{\rm ss}1}+\hat{A}_{{\rm ss}1}\hat{\alpha}_{21} \right)+\eta_2=0.
\end{align}
Equation \eqref{eq:SM-EOM-cavity-fields-symmetrized} can then be rearranged 
\begin{subequations}\label{eq:SM-cavity-fields-symmetrized-coupled-steady-state}
\begin{align}
\hat{A}_{{\rm ss}1}&=\frac{1}{4}\left[\left(\hat{\alpha}_{11}^{-1}\hat{\alpha}_{12}
+\hat{\alpha}_{12}\hat{\alpha}_{11}^{-1}\right)\hat{A}_{{\rm ss}2}
+\hat{A}_{{\rm ss}2}\left(\hat{\alpha}_{11}^{-1}\hat{\alpha}_{12}+\hat{\alpha}_{12}\hat{\alpha}_{11}^{-1}\right)\right]
+i\hat{\alpha}_{11}^{-1}\eta_1,
\label{eq:SM-cavity-fields-symmetrized-coupled-steady-state-A1}\\
\hat{A}_{{\rm ss}2}&=\frac{1}{4}\left[\left(\hat{\alpha}_{22}^{-1}\hat{\alpha}_{21}
+\hat{\alpha}_{21}\hat{\alpha}_{22}^{-1}\right)\hat{A}_{{\rm ss}1}
+\hat{A}_{{\rm ss}1}\left(\hat{\alpha}_{22}^{-1}\hat{\alpha}_{21}+\hat{\alpha}_{21}\hat{\alpha}_{22}^{-1}\right)\right]+
i\hat{\alpha}_{22}^{-1}\eta_2,
\label{eq:SM-cavity-fields-symmetrized-coupled-steady-state-A2}
\end{align}
\end{subequations}
where $\hat{\alpha}_{11}^{-1}$ and $\hat{\alpha}_{22}^{-1}$ are the inverse operators of 
$\hat{\alpha}_{11}$ and $\hat{\alpha}_{22}$, respectively, such that 
$\hat{\alpha}_{11}\hat{\alpha}_{11}^{-1}=\hat{\alpha}_{11}^{-1}\hat{\alpha}_{11}=\hat{1}$ and 
$\hat{\alpha}_{22}\hat{\alpha}_{22}^{-1}=\hat{\alpha}_{22}^{-1}\hat{\alpha}_{22}=\hat{1}$.
In order to make the subsequent analyses somewhat easier and trackable, 
we assume that all dual variables (except $\eta_j$ at this moment) are equal, 
namely, $\Delta_1=\Delta_2\equiv\Delta$, 
$\Delta_{\rm c1}=\Delta_{\rm c2}\equiv\Delta_{\rm c}$, and 
$\mathscr{G}_{ae}=\mathscr{G}_{be}\equiv \mathscr{G}_0$.
We also introduce $\tilde{\Delta}_{\rm c}\equiv\Delta_{\rm c}+i\kappa$ for a shorthand. 
We expand the inverse operators to the second order in a small unitless parameter 
$\xi\equiv2\mathscr{G}_0^2/\Delta\tilde{\Delta}_{\rm c}\ll1$ 
(and with $\langle \hat{N}_\tau\rangle\sim10^5$ one still has 
$\xi \langle \hat{N}_\tau\rangle\sim10^{-2}\ll1$, see Sec.~\ref{sec:model}
for more details),
\begin{align} \label{eq:SM-inverse-alpha11-22}
\hat{\alpha}_{11}^{-1}&=\left(\tilde{\Delta}_{\rm c}-\frac{2\mathscr{G}_0^2}{\Delta} \hat{N}_2 \right)^{-1} 
\simeq \tilde{\Delta}_{\rm c}^{-1} \left(1+\frac{2\mathscr{G}_0^2}{\Delta\tilde{\Delta}_{\rm c}} \hat{N}_2 
+\frac{4\mathscr{G}_0^4}{\Delta^2\tilde{\Delta}_{\rm c}^2} \hat{N}_2^2\right),
\nonumber\\
\hat{\alpha}_{22}^{-1}&=\left(\tilde{\Delta}_{\rm c}-\frac{2\mathscr{G}_0^2}{\Delta} \hat{N}_1 \right)^{-1} 
\simeq \tilde{\Delta}_{\rm c}^{-1} \left(1+\frac{2\mathscr{G}_0^2}{\Delta\tilde{\Delta}_{\rm c}} \hat{N}_1 
+\frac{4\mathscr{G}_0^4}{\Delta^2\tilde{\Delta}_{\rm c}^2} \hat{N}_1^2\right),
\end{align}
such that 
$\hat{\alpha}_{11}\hat{\alpha}_{11}^{-1}=\hat{\alpha}_{11}^{-1}\hat{\alpha}_{11}
=\hat{\alpha}_{22}\hat{\alpha}_{22}^{-1}=\hat{\alpha}_{22}^{-1}\hat{\alpha}_{22}
=\hat{1}+\mathcal{O}(\xi^3)$.
Note that the error in symmetrizing Eq.~\eqref{eq:SM-cavity-fields-symmetrized-coupled-steady-state} 
is also of order $\mathcal{O}(\xi^3)$. This can be easily checked by substituting, say, 
Eq.~\eqref{eq:SM-cavity-fields-symmetrized-coupled-steady-state-A1} in the first equation of 
\eqref{eq:SM-EOM-cavity-fields-symmetrized}. 
Equations \eqref{eq:SM-cavity-fields-symmetrized-coupled-steady-state-A1} and 
\eqref{eq:SM-cavity-fields-symmetrized-coupled-steady-state-A2} can now be simultaneously solved, yielding
\begin{align}\label{eq:a_1anda_2-general}
\hat{A}_{\rm ss1}&=i\hat{\Gamma}^{-1}
\left[\eta_1\hat{\alpha}_{11}^{-1}
+\frac{\eta_2}{4}\left(\hat{\alpha}_{11}^{-1}\hat{\alpha}_{12}\hat{\alpha}_{22}^{-1}
+\hat{\alpha}_{12}\hat{\alpha}_{11}^{-1}\hat{\alpha}_{22}^{-1}
+\hat{\alpha}_{22}^{-1}\hat{\alpha}_{11}^{-1}\hat{\alpha}_{12}
+\hat{\alpha}_{22}^{-1}\hat{\alpha}_{12}\hat{\alpha}_{11}^{-1} \right) \right],
\nonumber\\
\hat{A}_{\rm ss2}&=i\hat{\Gamma}^{-1}
\left[\eta_2\hat{\alpha}_{22}^{-1}
+\frac{\eta_1}{4}\left(\hat{\alpha}_{22}^{-1}\hat{\alpha}_{21}\hat{\alpha}_{11}^{-1}
+\hat{\alpha}_{21}\hat{\alpha}_{22}^{-1}\hat{\alpha}_{11}^{-1}
+\hat{\alpha}_{11}^{-1}\hat{\alpha}_{22}^{-1}\hat{\alpha}_{21} 
+\hat{\alpha}_{11}^{-1}\hat{\alpha}_{21}\hat{\alpha}_{22}^{-1}\right) \right],
\end{align}
where 
$\hat{\Gamma}=\left[1-
\frac{1}{2\tilde{\Delta}_{\rm c}^2}\left(\hat{\alpha}_{12}\hat{\alpha}_{21}
+\hat{\alpha}_{21}\hat{\alpha}_{12} \right) \right]$ up to $\xi^2$,
by noting $\hat{\alpha}_{12}=\hat{\alpha}_{21}^\dagger \propto \Omega'_R =2\mathcal{G}_0^2/\Delta$
and \eqref{eq:SM-inverse-alpha11-22}. We then have
\begin{align}\label{eq:SM-gamma-operator-inverse}
\hat{\Gamma}^{-1}&\simeq
1+\frac{1}{2\tilde{\Delta}_{\rm c}^2}\left(\hat{\alpha}_{12}\hat{\alpha}_{21}
+\hat{\alpha}_{21}\hat{\alpha}_{12} \right)=
1+\frac{2\mathscr{G}_0^4}{\Delta^2\tilde{\Delta}_{\rm c}^2}\left(\hat{S}_+\hat{S}_-+\hat{S}_-\hat{S}_+ \right),
\end{align}
up to $\mathcal{O}(\xi^3)$.
Using Eqs.~\eqref{eq:SM-alpha-matrix}, 
\eqref{eq:SM-inverse-alpha11-22}-\eqref{eq:SM-gamma-operator-inverse}, 
and retaining terms up to $\xi^2$, we obtain
\begin{align}\label{eq:SM-ss-solutions}
\hat{A}_{\rm ss1}&=\frac{i}{\tilde{\Delta}_{\rm c}}\left\{\eta_1
+ \frac{2\mathscr{G}_0^2}{\Delta\tilde{\Delta}_{\rm c}}\left(\eta_1\hat{N}_2+\eta_2\hat{S}_- \right)
+ \frac{4\mathscr{G}_0^4}{\Delta^2\tilde{\Delta}_{\rm c}^2} 
\left[\eta_1 \hat{N}_2^2+\frac{\eta_1}{2}\left(\hat{S}_+\hat{S}_-+\hat{S}_-\hat{S}_+ \right)
+\eta_2 \hat{N}\hat{S}_- \right] \right\},
\nonumber\\
\hat{A}_{\rm ss2}&=\frac{i}{\tilde{\Delta}_{\rm c}}\left\{\eta_2
+ \frac{2\mathscr{G}_0^2}{\Delta\tilde{\Delta}_{\rm c}}\left(\eta_2\hat{N}_1+\eta_1\hat{S}_+ \right)
+ \frac{4\mathscr{G}_0^4}{\Delta^2\tilde{\Delta}_{\rm c}^2} 
\left[\eta_2 \hat{N}_1^2+\frac{\eta_2}{2}\left(\hat{S}_+\hat{S}_-+\hat{S}_-\hat{S}_+ \right)
+\eta_1 \hat{N}\hat{S}_+ \right] \right\}.
\end{align}

By substituting steady-state solutions \eqref{eq:SM-ss-solutions} and their Hermitian conjugates 
in the Hamiltonian $H''^{(1)}_{\rm SO}$, Eq.~\eqref{eq:SM-2nd-UT-applied-H-integrated},
exercising symmetrization procedure again and retaining terms up to $\xi^2$, 
we can find an effective Hamiltonian which depends solely on the atomic operators. 
After some tedious though straightforward algebra, we obtain the cavity-field-eliminated effective 
many-body Hamiltonian 
\begin{align} \label{eq:app-cavity-field-eliminated-effective-many-particle-H}
H_{\rm eff}=
\int d^3r \left(\hat{\boldsymbol\Psi}^\dagger \mathcal{H}_{\rm SO}^{(1)} \hat{\boldsymbol\Psi}
+\frac {1}{2}g_{1} \hat{n}_{1}^2+\frac {1}{2}g_{2} \hat{n}_{2}^2+g_{12} \hat{n}_{1}\hat{n}_{2} \right)
+\sum_{\tau=1,2} U_{\tau} \hat{N}_\tau^2
+\left(U_{\pm}\hat{S}_+\hat{S}_- + U_{\mp}\hat{S}_-\hat{S}_+ \right)
+2U_{\rm ds} \hat{N} \hat{S}_x,                
\end{align}
where the cavity-field-eliminated, effective single-particle Hamiltonian density takes the familiar form
\begin{align} \label{eq:app-eff-1-particle-SOC-H}
\mathcal{H}_{\rm SO}^{(1)}=-\frac{\hbar^2}{2m}[\nabla_{\!\bot\!}^2-(-i\partial_{z}+k_R\sigma_z)^2]
+V_{\rm ext}(\mathbf r)+\frac{1}{2}\hbar\delta\sigma_z+\hbar\Omega_R \sigma_x,
\end{align}
with effective two-photon detuning and Raman coupling given by
\begin{gather} \label{eq:app-1-particle-coefficients} 
\delta\equiv\frac{2 \mathscr{G}_0^2 (\Delta_{\rm c}^2-\kappa^2)}
{\Delta(\Delta_{\rm c}^2+\kappa^2)^2}(\eta_2^2-\eta_1^2),
\nonumber\\
\Omega_R=
\frac{2\mathscr{G}_0^2}{\Delta(\Delta_{\rm c}^2+\kappa^2)^2}
\left(\Delta_{\rm c}^2-\kappa^2 - \frac{2\mathscr{G}_0^2 \Delta_{\rm c}}{\Delta}\right)
\eta_1 \eta_2=
\frac{\Omega'_R}{(\Delta_{\rm c}^2+\kappa^2)^2}
\left(\Delta_{\rm c}^2-\kappa^2 - \frac{2\mathscr{G}_0^2 \Delta_{\rm c}}{\Delta}\right)
\eta_1 \eta_2.
\end{gather}
(Note that $\delta'=0$, since we have assumed $\Delta_{\rm a1}=\Delta_{\rm a2}\equiv\Delta_{\rm a}$; 
cf.~Eqs.~\eqref{eq:SM-detunings} and \eqref{eq:SM-1st-adiabatic-pseudospin-energies}.) 
The coefficients of the cavity-mediated long-range interactions are found to be
\begin{align} \label{eq:SM-coefficients}
U_1=\frac{4\hbar \mathscr{G}_0^4 \Delta_{\rm c} (\Delta_{\rm c}^2-3\kappa^2)}
{\Delta^2(\Delta_{\rm c}^2+\kappa^2)^3}\eta_2^2,
\qquad
U_2=\frac{4\hbar \mathscr{G}_0^4 \Delta_{\rm c} (\Delta_{\rm c}^2-3\kappa^2)}
{\Delta^2(\Delta_{\rm c}^2+\kappa^2)^3}\eta_1^2,
\qquad
U_{\rm ds}=\frac{4\hbar \mathscr{G}_0^4 \Delta_{\rm c}\left(\Delta_{\rm c}^2-3\kappa^2 \right)}
{\Delta^2(\Delta_{\rm c}^2+\kappa^2)^3}\eta_1 \eta_2,\nonumber\\
U_{\pm}=\frac{4\hbar \mathscr{G}_0^4 \Delta_{\rm c}}{\Delta^2(\Delta_{\rm c}^2+\kappa^2)^3}
\left[\Delta_{\rm c}^2\eta_1^2-(\eta_1^2+2\eta_{2}^2)\kappa^2\right],
\qquad
U_{\mp}=\frac{4\hbar \mathscr{G}_0^4 \Delta_{\rm c}}{\Delta^2(\Delta_{\rm c}^2+\kappa^2)^3}
\left[\Delta_{\rm c}^2\eta_2^2-(\eta_2^2+2\eta_{1}^2)\kappa^2\right].
\end{align}
The terms with coefficients $U_{1/2}$, $U_{\pm/\mp}$, and $U_{\rm ds}$ in the effective
Hamiltonian \eqref{eq:app-cavity-field-eliminated-effective-many-particle-H}
are the cavity-mediated long-range interactions.
 Note that in the special case of $\eta_1=\eta_2\equiv\eta$, one has 
$\delta=0$ and $U_{1}=U_{2}=U_{\pm}=U_{\mp}=U_{\rm ds}\equiv U$.

\end{widetext} 


\bibliography{cbec5}

\begin{thebibliography}{35}
\bibitem[{Com()}]{Comment1}
\bibinfo{note}{The applied frequencies are assumed to be
  $\omega_1=\omega_2+\Delta\omega$ with $|\Delta\omega|/\omega_j\ll 1$. In the
  general case when $k_1\neq k_2$, one can define
  $k_c\equiv\frac{1}{2}(k_1+k_2)$ and $\Delta k_c\equiv k_1-k_2$. All the
  results still hold, except that $p_z$ is replaced by
  $p_z-\frac{1}{2}\hbar\Delta k_c$ after transferring to the co-moving frame of
  the cavity modes, which is just a Galilean transformation of the momentum.}
\end{thebibliography}

\end{document}